\documentstyle[12pt,epsfig,a4]{article}
\newcommand{\labe}[1]{\label{equ:#1}}
\newcommand{\labs}[1]{\label{sec:#1}}
\newcommand{\labf}[1]{\label{fig:#1}}
\newcommand{\labt}[1]{\label{tab:#1}}
\newcommand{\refe}[1]{\ref{equ:#1}}
\newcommand{\refs}[1]{\ref{sec:#1}}
\newcommand{\reff}[1]{\ref{fig:#1}}
\newcommand{\reft}[1]{\ref{tab:#1}}
\newcommand{\Ref}[1]{Ref.~\cite{#1}}

\newcommand{\Refss}[2]{Refs.~\cite{#1} and \cite{#2}}

\newcommand{\Eq}[1]{Eq.~(\refe{#1})}

\newcommand{\Figure}[1]{Figure~\reff{#1}}
\newcommand{\Fig}[1]{Fig.~\reff{#1}}

\newcommand{\Sec}[1]{Section~\refs{#1}}

\newcommand{\Table}[1]{Table~\reft{#1}}

\newcommand{\figcaption}[1]{\caption[.]{\small #1} \vspace{0.7cm}}
\newcommand{\tabcaption}[1]{\caption[.]{\small #1}}

\newcommand{\microns}{\mbox{$\mu$m}}
\newcommand{\GeVc}{\mbox{GeV$/c$}}
\newcommand{\GeVcc}{\mbox{GeV$/c^2$}}

\newcommand{\GeV}{{\rm GeV}}
\newcommand{\invps}{\mbox{ps$^{-1}$}}

\newcommand{\particle}[3]{\mbox{${\rm #1}_{\rm #2}^{\rm #3}$}}
\newcommand{\anti}[3]{\particle{\bar{#1}}{#2}{#3}}
\newcommand{\oscil}[3]{$\particle{#1}{#2}{#3}-\anti{#1}{#2}{#3}$}
\newcommand{\K}{\particle{K}{}{}}
\newcommand{\partitle}[3]{\particle{{\bf #1}}{\bf #2}{\bf #3}}

\newcommand{\Ds}{\particle{D}{s}{-}}

\newcommand{\Bs}{\particle{B}{s}{0}}
\newcommand{\Bsbar}{\anti{B}{s}{0}}
\newcommand{\Bd}{\particle{B}{d}{0}}

\newcommand{\Bu}{\particle{B}{}{+}}

\newcommand{\Bdu}{\Bd,\Bu}

\newcommand{\uds}{\particle{uds}{}{}}

\newcommand{\partsize}{\scriptsize}
\newcommand{\k}{{\partsize \K}}

\newcommand{\bs}{{\partsize \Bs}}
\newcommand{\bd}{{\partsize \Bd}}
\newcommand{\bu}{{\partsize \Bu}}

\newcommand{\decay}[2]{\mbox{$#1 \to #2$}}

\newcommand{\Z}[1]{\decay{\particle{Z}{}{}}{\particle{#1}{}{}\anti{#1}{}{}}}

\newcommand{\Rb}{\mbox{$R_{\rm b}$}}
\newcommand{\Rc}{\mbox{$R_{\rm c}$}}

\newcommand{\fs}{\mbox{$f_\bs$}}
\newcommand{\fd}{\mbox{$f_\bd$}}
\newcommand{\fu}{\mbox{$f_\bu$}}

\newcommand{\fbb}{\mbox{$f_{\mbox{\scriptsize \particle{b}{}{}-baryon}}$}}
\newcommand{\taus}{\mbox{$\tau_{\rm s}$}}

\newcommand{\dms}{\mbox{$\Delta m_{\rm s}$}}
\newcommand{\dmstrue}{\mbox{$\Delta m^{\rm true}_{\rm s}$}}

\newcommand{\dmd}{\mbox{$\Delta m_{\rm d}$}}
\newcommand{\dEdx}{\mbox{$dE/dx$}}
\newcommand{\chipi}{\mbox{$\chi_{\pi}$}}
\newcommand{\chiK}{\mbox{$\chi_{\k}$}}
\newcommand{\like}{\mbox{$\cal L$}}

\newcommand{\br}[1]{\mbox{$\cal B(#1)$}}
\newcommand{\BR}[2]{\br{\decay{#1}{#2}}}
\newcommand{\CL}[1]{#1\%~CL}
\newcommand{\dggs}{\mbox{$\Delta\Gamma_{\rm s}/\Gamma_{\rm s}$}}

\newcounter{pendingnumber}

\newcommand{\Res}{{\mathrm{Res}}}
\newcommand{\tO}{t_{0}}
\newcommand{\gO}{g_{0}}
\newcommand{\lO}{\l_{0}}
\newcommand{\muO}{\mu_0}
\newcommand{\ab}{\alpha\beta}
\newcommand{\sigl}{\sigma_l}
\newcommand{\Sldat}{S_l^{\mbox{\scriptsize dat}}}
\newcommand{\Sgdat}{S_g^{\mbox{\scriptsize dat}}}
\newcommand{\fldat}{f_l^{\mbox{\scriptsize dat}}}
\newcommand{\Sl}{S_l}

\newcommand{\eps}{\epsilon}
\newcommand{\sigAstat}{\mbox{$\sigma^{\mbox{\scriptsize stat}}_{\cal A}$}}
\newcommand{\sigAsyst}{\mbox{$\sigma^{\mbox{\scriptsize syst}}_{\cal A}$}}

\newcommand{\Dms}{\Delta m_{\mathrm s}}

\newcommand{\DGams}{\mbox{$\Delta\Gamma_{\mathrm s}$}}
\newcommand{\Gams}{\mbox{$\Gamma_{\mathrm s}$}}

\renewcommand{\thepage}{}

\begin{document}

\centerline{\large EUROPEAN LABORATORY FOR PARTICLE PHYSICS}
\vspace{15mm}
\begin{flushright}
CERN--EP/98--117\\
July 20, 1998 \\
\end{flushright}
 
\noindent
\vspace{30mm} \\
\centerline{\LARGE\bf  Search for  \partitle{B}{s}{0} oscillations }\vspace{2mm}\\
\centerline{\LARGE\bf using inclusive lepton events}
\vspace{4mm} \\
\begin{center}
{\em The ALEPH Collaboration
\footnote{See the following pages for the
     list of authors.}
}\\
\vspace{12mm}

\begin{abstract} \noindent
A  search for \Bs\ oscillations is performed using a sample of 
semileptonic \particle{b}{}{}-hadron decays collected by the ALEPH experiment
during 1991--1995.
Compared to previous inclusive lepton analyses, the proper time 
resolution and \particle{b}{}{}-flavour mistag rate are significantly improved. 
Additional sensitivity to \Bs\ mixing is obtained by identifying
subsamples of events having a \Bs\ purity which is higher  
than the average for the whole data sample. 
Unbinned maximum likelihood amplitude fits are performed to derive a
lower limit of  $\dms > 9.5~\invps$ at \CL{95}.
 Combining with the ALEPH
\Ds\ based analyses yields $\dms > 9.6~\invps$ at \CL{95}.
\end{abstract}

\vfill 
 {\em (Submitted to European Physical Journal C) } \vspace{15mm}
\end{center}

\pagestyle{empty}
\newpage
\small
%
\newlength{\saveparskip}
\newlength{\savetextheight}
\newlength{\savetopmargin}
\newlength{\savetextwidth}
\newlength{\saveoddsidemargin}
\newlength{\savetopsep}
\setlength{\saveparskip}{\parskip}
\setlength{\savetextheight}{\textheight}
\setlength{\savetopmargin}{\topmargin}
\setlength{\savetextwidth}{\textwidth}
\setlength{\saveoddsidemargin}{\oddsidemargin}
\setlength{\savetopsep}{\topsep}
%
%
\setlength{\parskip}{0.0cm}
\setlength{\textheight}{25.0cm}
\setlength{\topmargin}{-1.5cm}
\setlength{\textwidth}{16 cm}
\setlength{\oddsidemargin}{-0.0cm}
\setlength{\topsep}{1mm}
\pretolerance=10000
\centerline{\large\bf The ALEPH Collaboration}
\footnotesize
\vspace{0.5cm}
{\raggedbottom
\begin{sloppypar}
\samepage\noindent
R.~Barate,
D.~Buskulic,
D.~Decamp,
P.~Ghez,
C.~Goy,
\mbox{J.-P.~Lees},
A.~Lucotte,
E.~Merle,
\mbox{M.-N.~Minard},
\mbox{J.-Y.~Nief},
B.~Pietrzyk
\nopagebreak
\begin{center}
\parbox{15.5cm}{\sl\samepage
Laboratoire de Physique des Particules (LAPP), IN$^{2}$P$^{3}$-CNRS,
F-74019 Annecy-le-Vieux Cedex, France}
\end{center}\end{sloppypar}
\vspace{2mm}
\begin{sloppypar}
\noindent
R.~Alemany,
G.~Boix,
M.P.~Casado,
M.~Chmeissani,
J.M.~Crespo,
M.~Delfino,
E.~Fernandez,
\mbox{M.~Fernandez-Bosman},
Ll.~Garrido,$^{15}$
E.~Graug\`{e}s,
A.~Juste,
M.~Martinez,
G.~Merino,
R.~Miquel,
Ll.M.~Mir,
I.C.~Park,
A.~Pascual,
I.~Riu,
F.~Sanchez
\nopagebreak
\begin{center}
\parbox{15.5cm}{\sl\samepage
Institut de F\'{i}sica d'Altes Energies, Universitat Aut\`{o}noma
de Barcelona, E-08193 Bellaterra (Barcelona), Spain$^{7}$}
\end{center}\end{sloppypar}
\vspace{2mm}
\begin{sloppypar}
\noindent
A.~Colaleo,
D.~Creanza,
M.~de~Palma,
G.~Gelao,
G.~Iaselli,
G.~Maggi,
M.~Maggi,
S.~Nuzzo,
A.~Ranieri,
G.~Raso,
F.~Ruggieri,
G.~Selvaggi,
L.~Silvestris,
P.~Tempesta,
A.~Tricomi,$^{3}$
G.~Zito
\nopagebreak
\begin{center}
\parbox{15.5cm}{\sl\samepage
Dipartimento di Fisica, INFN Sezione di Bari, I-70126
Bari, Italy}
\end{center}\end{sloppypar}
\vspace{2mm}
\begin{sloppypar}
\noindent
X.~Huang,
J.~Lin,
Q. Ouyang,
T.~Wang,
Y.~Xie,
R.~Xu,
S.~Xue,
J.~Zhang,
L.~Zhang,
W.~Zhao
\nopagebreak
\begin{center}
\parbox{15.5cm}{\sl\samepage
Institute of High-Energy Physics, Academia Sinica, Beijing, The People's
Republic of China$^{8}$}
\end{center}\end{sloppypar}
\vspace{2mm}
\begin{sloppypar}
\noindent
D.~Abbaneo,
U.~Becker,
\mbox{P.~Bright-Thomas},$^{24}$
D.~Casper,
M.~Cattaneo,
F.~Cerutti,
V.~Ciulli,
G.~Dissertori,
H.~Drevermann,
R.W.~Forty,
M.~Frank,
R.~Hagelberg,
A.W. Halley,
J.B.~Hansen,
J.~Harvey,
P.~Janot,
B.~Jost,
I.~Lehraus,
P.~Mato,
A.~Minten,
L.~Moneta,$^{21}$
A.~Pacheco,
F.~Ranjard,
L.~Rolandi,
D.~Rousseau,
D.~Schlatter,
M.~Schmitt,$^{20}$
O.~Schneider,
W.~Tejessy,
F.~Teubert,
I.R.~Tomalin,
H.~Wachsmuth
\nopagebreak
\begin{center}
\parbox{15.5cm}{\sl\samepage
European Laboratory for Particle Physics (CERN), CH-1211 Geneva 23,
Switzerland}
\end{center}\end{sloppypar}
\vspace{2mm}
\begin{sloppypar}
\noindent
Z.~Ajaltouni,
F.~Badaud,
G.~Chazelle,
O.~Deschamps,
A.~Falvard,
C.~Ferdi,
P.~Gay,
C.~Guicheney,
P.~Henrard,
J.~Jousset,
B.~Michel,
S.~Monteil,
\mbox{J-C.~Montret},
D.~Pallin,
P.~Perret,
F.~Podlyski,
J.~Proriol,
P.~Rosnet
\nopagebreak
\begin{center}
\parbox{15.5cm}{\sl\samepage
Laboratoire de Physique Corpusculaire, Universit\'e Blaise Pascal,
IN$^{2}$P$^{3}$-CNRS, Clermont-Ferrand, F-63177 Aubi\`{e}re, France}
\end{center}\end{sloppypar}
\vspace{2mm}
\begin{sloppypar}
\noindent
J.D.~Hansen,
J.R.~Hansen,
P.H.~Hansen,
B.S.~Nilsson,
B.~Rensch,
A.~W\"a\"an\"anen
\begin{center}
\parbox{15.5cm}{\sl\samepage
Niels Bohr Institute, DK-2100 Copenhagen, Denmark$^{9}$}
\end{center}\end{sloppypar}
\vspace{2mm}
\begin{sloppypar}
\noindent
G.~Daskalakis,
A.~Kyriakis,
C.~Markou,
E.~Simopoulou,
I.~Siotis,
A.~Vayaki
\nopagebreak
\begin{center}
\parbox{15.5cm}{\sl\samepage
Nuclear Research Center Demokritos (NRCD), GR-15310 Attiki, Greece}
\end{center}\end{sloppypar}
\vspace{2mm}
\begin{sloppypar}
\noindent
A.~Blondel,
G.~Bonneaud,
\mbox{J.-C.~Brient},
P.~Bourdon,
A.~Roug\'{e},
M.~Rumpf,
A.~Valassi,$^{6}$
M.~Verderi,
H.~Videau
\nopagebreak
\begin{center}
\parbox{15.5cm}{\sl\samepage
Laboratoire de Physique Nucl\'eaire et des Hautes Energies, Ecole
Polytechnique, IN$^{2}$P$^{3}$-CNRS, \mbox{F-91128} Palaiseau Cedex, France}
\end{center}\end{sloppypar}
\vspace{2mm}
\begin{sloppypar}
\noindent
E.~Focardi,
G.~Parrini,
K.~Zachariadou
\nopagebreak
\begin{center}
\parbox{15.5cm}{\sl\samepage
Dipartimento di Fisica, Universit\`a di Firenze, INFN Sezione di Firenze,
I-50125 Firenze, Italy}
\end{center}\end{sloppypar}
\vspace{2mm}
\begin{sloppypar}
\noindent
M.~Corden,
C.~Georgiopoulos,
D.E.~Jaffe
\nopagebreak
\begin{center}
\parbox{15.5cm}{\sl\samepage
Supercomputer Computations Research Institute,
Florida State University,
Tallahassee, FL 32306-4052, USA $^{13,14}$}
\end{center}\end{sloppypar}
\vspace{2mm}
\begin{sloppypar}
\noindent
A.~Antonelli,
G.~Bencivenni,
G.~Bologna,$^{4}$
F.~Bossi,
P.~Campana,
G.~Capon,
V.~Chiarella,
G.~Felici,
P.~Laurelli,
G.~Mannocchi,$^{5}$
F.~Murtas,
G.P.~Murtas,
L.~Passalacqua,
\mbox{M.~Pepe-Altarelli}
\nopagebreak
\begin{center}
\parbox{15.5cm}{\sl\samepage
Laboratori Nazionali dell'INFN (LNF-INFN), I-00044 Frascati, Italy}
\end{center}\end{sloppypar}
\vspace{2mm}
\begin{sloppypar}
\noindent
L.~Curtis,
J.G.~Lynch,
P.~Negus,
V.~O'Shea,
C.~Raine,
J.M.~Scarr,
K.~Smith,
\mbox{P.~Teixeira-Dias},
A.S.~Thompson,
E.~Thomson
\nopagebreak
\begin{center}
\parbox{15.5cm}{\sl\samepage
Department of Physics and Astronomy, University of Glasgow, Glasgow G12
8QQ,United Kingdom$^{10}$}
\end{center}\end{sloppypar}
\pagebreak
\vspace{2mm}
\begin{sloppypar}
\noindent
O.~Buchm\"uller,
S.~Dhamotharan,
C.~Geweniger,
G.~Graefe,
P.~Hanke,
G.~Hansper,
V.~Hepp,
E.E.~Kluge,
A.~Putzer,
J.~Sommer,
K.~Tittel,
S.~Werner,
M.~Wunsch
\nopagebreak
\begin{center}
\parbox{15.5cm}{\sl\samepage
Institut f\"ur Hochenergiephysik, Universit\"at Heidelberg, D-69120
Heidelberg, Germany$^{16}$}
\end{center}\end{sloppypar}
\vspace{2mm}
\begin{sloppypar}
\noindent
R.~Beuselinck,
D.M.~Binnie,
W.~Cameron,
P.J.~Dornan,$^{2}$
M.~Girone,
S.~Goodsir,
E.B.~Martin,
N.~Marinelli,
A.~Moutoussi,
J.~Nash,
J.K.~Sedgbeer,
P.~Spagnolo,
M.D.~Williams
\nopagebreak
\begin{center}
\parbox{15.5cm}{\sl\samepage
Department of Physics, Imperial College, London SW7 2BZ,
United Kingdom$^{10}$}
\end{center}\end{sloppypar}
\vspace{2mm}
\begin{sloppypar}
\noindent
V.M.~Ghete,
P.~Girtler,
E.~Kneringer,
D.~Kuhn,
G.~Rudolph
\nopagebreak
\begin{center}
\parbox{15.5cm}{\sl\samepage
Institut f\"ur Experimentalphysik, Universit\"at Innsbruck, A-6020
Innsbruck, Austria$^{18}$}
\end{center}\end{sloppypar}
\vspace{2mm}
\begin{sloppypar}
\noindent
A.P.~Betteridge,
C.K.~Bowdery,
P.G.~Buck,
P.~Colrain,
G.~Crawford,
A.J.~Finch,
F.~Foster,
G.~Hughes,
R.W.L.~Jones,
N.A.~Robertson,
M.I.~Williams
\nopagebreak
\begin{center}
\parbox{15.5cm}{\sl\samepage
Department of Physics, University of Lancaster, Lancaster LA1 4YB,
United Kingdom$^{10}$}
\end{center}\end{sloppypar}
\vspace{2mm}
\begin{sloppypar}
\noindent
I.~Giehl,
C.~Hoffmann,
K.~Jakobs,
K.~Kleinknecht,
G.~Quast,
B.~Renk,
E.~Rohne,
\mbox{H.-G.~Sander},
P.~van~Gemmeren,
C.~Zeitnitz
\nopagebreak
\begin{center}
\parbox{15.5cm}{\sl\samepage
Institut f\"ur Physik, Universit\"at Mainz, D-55099 Mainz, Germany$^{16}$}
\end{center}\end{sloppypar}
\vspace{2mm}
\begin{sloppypar}
\noindent
J.J.~Aubert,
C.~Benchouk,
A.~Bonissent,
G.~Bujosa,
J.~Carr,$^{2}$
P.~Coyle,
F.~Etienne,
O.~Leroy,
F.~Motsch,
P.~Payre,
M.~Talby,
A.~Sadouki,
M.~Thulasidas,
K.~Trabelsi
\nopagebreak
\begin{center}
\parbox{15.5cm}{\sl\samepage
Centre de Physique des Particules, Facult\'e des Sciences de Luminy,
IN$^{2}$P$^{3}$-CNRS, F-13288 Marseille, France}
\end{center}\end{sloppypar}
\vspace{2mm}
\begin{sloppypar}
\noindent
M.~Aleppo,
M.~Antonelli,
F.~Ragusa
\nopagebreak
\begin{center}
\parbox{15.5cm}{\sl\samepage
Dipartimento di Fisica, Universit\`a di Milano e INFN Sezione di Milano,
I-20133 Milano, Italy}
\end{center}\end{sloppypar}
\vspace{2mm}
\begin{sloppypar}
\noindent
R.~Berlich,
V.~B\"uscher,
G.~Cowan,
H.~Dietl,
G.~Ganis,
G.~L\"utjens,
C.~Mannert,
W.~M\"anner,
\mbox{H.-G.~Moser},
S.~Schael,
R.~Settles,
H.~Seywerd,
H.~Stenzel,
W.~Wiedenmann,
G.~Wolf
\nopagebreak
\begin{center}
\parbox{15.5cm}{\sl\samepage
Max-Planck-Institut f\"ur Physik, Werner-Heisenberg-Institut,
D-80805 M\"unchen, Germany\footnotemark[16]}
\end{center}\end{sloppypar}
\vspace{2mm}
\begin{sloppypar}
\noindent
J.~Boucrot,
O.~Callot,
S.~Chen,
A.~Cordier,
M.~Davier,
L.~Duflot,
\mbox{J.-F.~Grivaz},
Ph.~Heusse,
A.~H\"ocker,
A.~Jacholkowska,
D.W.~Kim,$^{12}$
F.~Le~Diberder,
J.~Lefran\c{c}ois,
\mbox{A.-M.~Lutz},
\mbox{M.-H.~Schune},
E.~Tournefier,
\mbox{J.-J.~Veillet},
I.~Videau,
D.~Zerwas
\nopagebreak
\begin{center}
\parbox{15.5cm}{\sl\samepage
Laboratoire de l'Acc\'el\'erateur Lin\'eaire, Universit\'e de Paris-Sud,
IN$^{2}$P$^{3}$-CNRS, F-91898 Orsay Cedex, France}
\end{center}\end{sloppypar}
\vspace{2mm}
\begin{sloppypar}
\noindent
\samepage
P.~Azzurri,
G.~Bagliesi,$^{2}$
G.~Batignani,
S.~Bettarini,
T.~Boccali,
C.~Bozzi,
G.~Calderini,
M.~Carpinelli,
M.A.~Ciocci,
R.~Dell'Orso,
R.~Fantechi,
I.~Ferrante,
L.~Fo\`{a},$^{1}$
F.~Forti,
A.~Giassi,
M.A.~Giorgi,
A.~Gregorio,
F.~Ligabue,
A.~Lusiani,
P.S.~Marrocchesi,
A.~Messineo,
F.~Palla,
G.~Rizzo,
G.~Sanguinetti,
A.~Sciab\`a,
G.~Sguazzoni,
R.~Tenchini,
G.~Tonelli,$^{19}$
C.~Vannini,
A.~Venturi,
P.G.~Verdini
\samepage
\begin{center}
\parbox{15.5cm}{\sl\samepage
Dipartimento di Fisica dell'Universit\`a, INFN Sezione di Pisa,
e Scuola Normale Superiore, I-56010 Pisa, Italy}
\end{center}\end{sloppypar}
\vspace{2mm}
\begin{sloppypar}
\noindent
G.A.~Blair,
L.M.~Bryant,
J.T.~Chambers,
M.G.~Green,
T.~Medcalf,
P.~Perrodo,
J.A.~Strong,
\mbox{J.H.~von~Wimmersperg-Toeller}
\nopagebreak
\begin{center}
\parbox{15.5cm}{\sl\samepage
Department of Physics, Royal Holloway \& Bedford New College,
University of London, Surrey TW20 OEX, United Kingdom$^{10}$}
\end{center}\end{sloppypar}
\vspace{2mm}
\begin{sloppypar}
\noindent
D.R.~Botterill,
R.W.~Clifft,
T.R.~Edgecock,
P.R.~Norton,
J.C.~Thompson,
A.E.~Wright
\nopagebreak
\begin{center}
\parbox{15.5cm}{\sl\samepage
Particle Physics Dept., Rutherford Appleton Laboratory,
Chilton, Didcot, Oxon OX11 OQX, United Kingdom$^{10}$}
\end{center}\end{sloppypar}
\vspace{2mm}
\begin{sloppypar}
\noindent
\mbox{B.~Bloch-Devaux},
P.~Colas,
S.~Emery,
W.~Kozanecki,
E.~Lan\c{c}on,$^{2}$
\mbox{M.-C.~Lemaire},
E.~Locci,
P.~Perez,
J.~Rander,
\mbox{J.-F.~Renardy},
A.~Roussarie,
\mbox{J.-P.~Schuller},
J.~Schwindling,
A.~Trabelsi,
B.~Vallage
\nopagebreak
\begin{center}
\parbox{15.5cm}{\sl\samepage
CEA, DAPNIA/Service de Physique des Particules,
CE-Saclay, F-91191 Gif-sur-Yvette Cedex, France$^{17}$}
\end{center}\end{sloppypar}
\pagebreak
\vspace{2mm}
\begin{sloppypar}
\noindent
S.N.~Black,
J.H.~Dann,
R.P.~Johnson,
H.Y.~Kim,
N.~Konstantinidis,
A.M.~Litke,
M.A.~McNeil,
G.~Taylor
\nopagebreak
\vspace{-2.7ex}
\begin{center}
\parbox{15.5cm}{\sl\samepage
Institute for Particle Physics, University of California at
Santa Cruz, Santa Cruz, CA 95064, USA$^{22}$}
\end{center}\end{sloppypar}
\vspace{2mm}
\begin{sloppypar}
\noindent
C.N.~Booth,
S.~Cartwright,
F.~Combley,
M.S.~Kelly,
M.~Lehto,
L.F.~Thompson
\nopagebreak
\begin{center}
\parbox{15.5cm}{\sl\samepage
Department of Physics, University of Sheffield, Sheffield S3 7RH,
United Kingdom$^{10}$}
\end{center}\end{sloppypar}
\vspace{2mm}
\begin{sloppypar}
\noindent
K.~Affholderbach,
A.~B\"ohrer,
S.~Brandt,
C.~Grupen,
P.~Saraiva,
L.~Smolik,
F.~Stephan
\nopagebreak
\begin{center}
\parbox{15.5cm}{\sl\samepage
Fachbereich Physik, Universit\"at Siegen, D-57068 Siegen,
 Germany$^{16}$}
\end{center}\end{sloppypar}
\vspace{2mm}
\begin{sloppypar}
\noindent
G.~Giannini,
B.~Gobbo,
G.~Musolino
\nopagebreak
\begin{center}
\parbox{15.5cm}{\sl\samepage
Dipartimento di Fisica, Universit\`a di Trieste e INFN Sezione di Trieste,
I-34127 Trieste, Italy}
\end{center}\end{sloppypar}
\vspace{2mm}
\begin{sloppypar}
\noindent
J.~Rothberg,
S.~Wasserbaech
\nopagebreak
\begin{center}
\parbox{15.5cm}{\sl\samepage
Experimental Elementary Particle Physics, University of Washington, WA 98195
Seattle, U.S.A.}
\end{center}\end{sloppypar}
\vspace{2mm}
\begin{sloppypar}
\noindent
S.R.~Armstrong,
E.~Charles,
P.~Elmer,
D.P.S.~Ferguson,
Y.~Gao,
S.~Gonz\'{a}lez,
T.C.~Greening,
O.J.~Hayes,
H.~Hu,
S.~Jin,
P.A.~McNamara III,
J.M.~Nachtman,$^{23}$
J.~Nielsen,
W.~Orejudos,
Y.B.~Pan,
Y.~Saadi,
I.J.~Scott,
J.~Walsh,
Sau~Lan~Wu,
X.~Wu,
G.~Zobernig
\nopagebreak
\begin{center}
\parbox{15.5cm}{\sl\samepage
Department of Physics, University of Wisconsin, Madison, WI 53706,
USA$^{11}$}
\end{center}\end{sloppypar}
}
\footnotetext[1]{Now at CERN, 1211 Geneva 23,
Switzerland.}
\footnotetext[2]{Also at CERN, 1211 Geneva 23, Switzerland.}
\footnotetext[3]{Also at Dipartimento di Fisica, INFN, Sezione di Catania, Catania, Italy.}
\footnotetext[4]{Also Istituto di Fisica Generale, Universit\`{a} di
Torino, Torino, Italy.}
\footnotetext[5]{Also Istituto di Cosmo-Geofisica del C.N.R., Torino,
Italy.}
\footnotetext[6]{Supported by the Commission of the European Communities,
contract ERBCHBICT941234.}
\footnotetext[7]{Supported by CICYT, Spain.}
\footnotetext[8]{Supported by the National Science Foundation of China.}
\footnotetext[9]{Supported by the Danish Natural Science Research Council.}
\footnotetext[10]{Supported by the UK Particle Physics and Astronomy Research
Council.}
\footnotetext[11]{Supported by the US Department of Energy, grant
DE-FG0295-ER40896.}
\footnotetext[12]{Permanent address: Kangnung National University, Kangnung, 
Korea.}
\footnotetext[13]{Supported by the US Department of Energy, contract
DE-FG05-92ER40742.}
\footnotetext[14]{Supported by the US Department of Energy, contract
DE-FC05-85ER250000.}
\footnotetext[15]{Permanent address: Universitat de Barcelona, 08208 Barcelona,
Spain.}
\footnotetext[16]{Supported by the Bundesministerium f\"ur Bildung,
Wissenschaft, Forschung und Technologie, Germany.}
\footnotetext[17]{Supported by the Direction des Sciences de la
Mati\`ere, C.E.A.}
\footnotetext[18]{Supported by Fonds zur F\"orderung der wissenschaftlichen
Forschung, Austria.}
\footnotetext[19]{Also at Istituto di Matematica e Fisica,
Universit\`a di Sassari, Sassari, Italy.}
\footnotetext[20]{Now at Harvard University, Cambridge, MA 02138, U.S.A.}
\footnotetext[21]{Now at University of Geneva, 1211 Geneva 4, Switzerland.}
\footnotetext[22]{Supported by the US Department of Energy,
grant DE-FG03-92ER40689.}
\footnotetext[23]{Now at University of California at Los Angeles (UCLA),
Los Angeles, CA 90024, U.S.A.}
\footnotetext[24]{Now at School of Physics and Astronomy,
Birmingham B15 2TT, U.K.}
%
%
\setlength{\parskip}{\saveparskip}
\setlength{\textheight}{\savetextheight}
\setlength{\topmargin}{\savetopmargin}
\setlength{\textwidth}{\savetextwidth}
\setlength{\oddsidemargin}{\saveoddsidemargin}
\setlength{\topsep}{\savetopsep}
\normalsize
\newpage
\pagestyle{plain}
\setcounter{page}{1}

\newpage \pagestyle{plain} \setcounter{page}{1}

\newpage 
\renewcommand{\thepage}{\arabic{page}}
\setcounter{page}{1}
\pagestyle{plain}

\section{Introduction}

Flavour non-conservation in charged weak current interactions allows
mixing between the \Bs\ and \Bsbar\ flavour states.
The proper-time probability density function of a \Bs\ meson
which is known to have mixed
oscillates. 
The oscillation frequency is proportional to \dms, the
mass difference between the mass eigenstates.
Within the framework of the Standard Model, a measurement of 
the ratio  \dms/\dmd\ 
(\dmd\ being the mass difference in the \oscil{B}{d}{0} system)
would allow the extraction of the ratio of 
Cabibbo-Kobayashi-Maskawa (CKM) quark mixing matrix elements
$|V_{\rm ts}/V_{\rm td}|$.

Although the slower \Bd\ oscillations are now well established,
the faster \Bs\ oscillations remain to be detected. Previous 
ALEPH analyses searching for \Bs\ oscillations have either been based 
on semi-exclusive selections in which a \Ds\ is fully 
reconstructed~\cite{ALEPH-DS-LEPTON,ALEPH-DSHAD} or on more inclusive lepton 
selections~\cite{ALEPH-DILEPTON,ALEPH-LEPTON-JET-WISCONSIN,
ALEPH-WARSAW-COMBINATION}. 
Although the latter suffer from a lower \Bs\ purity and poorer 
proper time resolution they have the advantage of larger statistics.   

The analysis presented here is also based on an inclusive lepton sample. 
Compared to the previous ALEPH inclusive lepton 
analysis~\cite{ALEPH-LEPTON-JET-WISCONSIN}, the following 
improvements are made to increase the sensitivity to \Bs\ mixing.

\begin{itemize}

\item {\bf Decay length resolution:} 
An improved decay length resolution is 
obtained by applying tight selection cuts to remove events likely to 
have misassigned tracks between the primary and the \Bs\ vertex. 
In addition an estimate of the decay length uncertainty is used on an event-by-event basis, 
rather than assuming the same average decay length uncertainty for all events,
as used in previous analyses. 

\item{\bf Boost resolution:} 
A nucleated jet algorithm is used for an improved estimate of
the momentum of the \particle{b}{}{}-hadrons. 

\item{\bf \boldmath \Bs\ purity classes:}
Various properties of the events, such as the charge of the reconstructed 
\particle{b}{}{}-hadron vertex and the presence of kaons are used to enhance
the fraction of \Bs\ in subsamples of the data. 

\item {\bf Initial and final state tagging:} 
The \particle{b}{}{}-flavour tagging method previously used for the \Ds\ based analyses 
\cite{ALEPH-DS-LEPTON,ALEPH-DSHAD} is applied.
In this method discriminating variables are used to construct 
mistag probabilities and sample composition fractions 
estimated on an event-by-event basis. 
As a result, all events are tagged and the effective mistag rate is reduced.

\end{itemize} 

This paper details these improvements and is organized as follows. 
After a brief description of the ALEPH detector, the event 
selection is described in \Sec{eventsel} and the \Bs\ purity classification 
procedure in \Sec{enrichment}. The next two sections explain 
the proper time reconstruction and the procedure for tagging the initial and 
final state b quark charge. 
The likelihood function is presented in \Sec{likelihood}
and the $\Dms$ results in \Sec{results}.
In \Sec{sec_syst}  the systematic uncertainties 
are described, and in \Sec{sec_checks} additional checks of the 
analysis presented. Finally the combination of this analysis 
with the ALEPH \Ds\ based analyses is described in \Sec{combination}.

\section{The ALEPH detector} \labs{detector}
The ALEPH detector and its performance from 1991 to 1995
are described in detail elsewhere~\cite{ALEPH-DETECTOR,ALEPH-PERFORMANCE},
and only a brief
overview of the apparatus is given here. Surrounding the beam pipe, 
a high resolution vertex detector (VDET) consists of two layers of
double-sided silicon microstrip detectors,
positioned at average radii of 6.5~cm and 11.3~cm,
and covering 85\% and 69\% of the solid angle respectively.
The spatial resolution for the $r\phi$ and
$z$ projections (transverse to and along the beam axis, respectively) is
12~\microns\ at normal incidence. The vertex detector is surrounded
by a drift chamber with eight coaxial wire layers with an outer
radius of 26~cm and by a time projection chamber
(TPC) that measures up to 21~three-dimensional points per track at radii
between 30~cm and 180~cm. These detectors are immersed
in an axial magnetic field of 1.5~T and together measure the momenta of
charged particles with a resolution
$\sigma (p)/p = 6 \times 10^{-4} \, p_{\mathrm T}
                   \oplus 0.005$ ($p_{\mathrm T}$ in \GeVc).
The resolution of the three-dimensional impact parameter in the
transverse and longitudinal view,
for tracks having information from all tracking detectors and two
VDET hits (a VDET ``hit'' being defined as having information 
from both $r\phi$ and $z$ views), can be parametrized as
$\sigma = 25\, \mu{\mathrm m} + 95\, \mu{\mathrm m}/p$ ($p$ in \GeVc).
The TPC also provides up to 338~measurements of the specific ionization
of a charged particle. In the following, the \dEdx\ information is
considered as available if more than 50 samples are present.
Particle identification is based on the \dEdx\ estimator \chipi\ (\chiK),
defined as the difference between the measured and expected ionization
expressed in terms of standard deviations for the $\pi$ (\K) mass hypothesis.
The TPC is surrounded by a lead/proportional-chamber electromagnetic
calorimeter segmented into $0.9^{\circ} \times 0.9^{\circ}$ projective towers
and read out in three sections in depth, with energy resolution
$\sigma (E)/E = 0.18/\sqrt{E} + 0.009 $ ($E$ in GeV). The iron return
yoke of the magnet is instrumented with streamer tubes to
form a hadron calorimeter, with a thickness of over 7
interaction lengths and is surrounded by two additional double-layers
of streamer tubes to aid muon identification.
An algorithm combines all these measurements to provide a determination
of the energy flow~\cite{ALEPH-PERFORMANCE} with an uncertainty
on the total measurable energy of
\mbox{$\sigma(E) = (0.6\sqrt{E/{\mathrm {GeV}}} + 0.6)~\GeV$.}

\section{Event selection} \labs{eventsel}

This analysis uses approximately 4 million hadronic \particle{Z}{}{} events 
recorded by the ALEPH detector from 1991 to 1995 at centre of mass energies 
close to the \particle{Z}{}{} peak and selected
with the charged particle requirements described in \Ref{ALEPH-HADRONIC}.
It relies on Monte Carlo samples of fully simulated \Z{q} events.
The Monte Carlo generator is based 
on JETSET 7.4~\cite{LUND} with updated branching ratios for heavy flavour
decays. Monte Carlo events are reweighted to the 
physics parameters listed in \Table{phyparams}.

Events for which the cosine of the angle between the thrust 
axis and the beam axis is less than 0.85 are selected.
Using the plane perpendicular to the thrust axis, 
the event is split into two hemispheres. 
Electrons and muons are identified using
the standard ALEPH lepton selection criteria~\cite{lepton-ID}.
Events containing at least one such lepton with
momentum above 3~\GeVc\ are kept. 
The leptons are then associated to their closest jet (constructed using the 
JADE algorithm~\cite{JADE} with $y_{\mbox{\scriptsize cut}}=0.004$) 
and a transverse momentum  $p_T$ with respect to the jet is calculated 
with the lepton momentum removed from the jet.
 Only leptons with $p_T >1.25$~\GeVc~are selected.
In the case that more than one lepton in an event satisfies this 
requirement, only the lepton with the highest momentum is used as a
candidate for a \Bs\ decay product.

The \particle{e}{}{+}\particle{e}{}{-} interaction point is
reconstructed on an event-by-event basis using the constraint of the
average beam spot position and envelope~\cite{1993_Rb_paper}.

A charm vertex is then reconstructed in the lepton hemisphere using the
algorithm described in~\Ref{ALEPH-DILEPTON}. Charged particles
in this hemisphere (other than the selected lepton) 
are assigned to either the interaction point or a single 
displaced secondary vertex. A three-dimensional grid search is performed for
the secondary vertex position to find the combination of
assignments that has the greatest reduction in $\chi^2$ 
as compared to the case when all tracks are assumed to come from the 
interaction point. Tracks are required to come within $3\sigma$ of their 
assigned vertex. The position resolution of this ``charm vertex'' is subsequently improved 
by removing those tracks having a momentum below 1.5~\GeVc\ or an impact parameter 
significance relative to the charm vertex larger than $1.4\sigma$.  The remaining tracks are 
then re-vertexed to form the reconstructed ``charm particle''. 
If only one track passes the requirements, it 
serves as the charm particle. The event is rejected
if no track remains, or none of the tracks assigned to the charm vertex
have at least one vertex detector hit. 
The charm particle is
then combined with the lepton to form a candidate \particle{b}{}{}-hadron vertex. The 
lepton is required to have at least one vertex detector hit and the $\chi^2$ per 
degree of freedom of the reconstructed \particle{b}{}{}-hadron
vertex is required to be less than 25. 

The energy $E_{\mathrm c}$ of the charm particle is estimated by clustering 
a jet, using the JADE algorithm, around the charged tracks at the charm 
vertex until a mass of $2.7~\GeVcc$ is reached. 
To reduce the influence of fragmentation particles on the estimate of $E_{\mathrm c}$, 
charged and neutral particles with energies less than 
0.5~GeV are excluded from the clustering~\cite{DR}.  
The neutrino energy $E_{\nu}$ is estimated from the missing energy in the 
lepton hemisphere taking into account the 
measured mass in each hemisphere~\cite{Bs_lifetime}. 
Assuming the direction of flight of the \particle{b}{}{}-hadron to be
that of its associated jet,
an estimate of the \particle{b}{}{}-hadron mass can be calculated 
from the energy of the neutrino and the four-vectors of the 
charm particle and the lepton. 

\begin{table}
\tabcaption{Values of the physics parameters assumed in this analysis.} 
\vskip 0.1cm
\begin{center}
\begin{tabular}{|c|c|c| }
\hline
Physics parameter & Value and uncertainty & Reference \\
\hline \hline
\Bs\   lifetime   & $1.49  \pm 0.06 $~ps & \cite{Schneider} \\
\Bd\   lifetime   & $1.57  \pm 0.04 $~ps & \cite{Schneider} \\
\Bu\   lifetime   & $1.67  \pm 0.04 $~ps & \cite{Schneider} \\
\particle{b}{}{}-baryon lifetime   & $1.22  \pm 0.05 $~ps & \cite{Schneider} \\
\dmd           & $0.463 \pm 0.018~\invps$ & \cite{Schneider} \\ 
\hline \rule{0pt}{11pt} 
$\Rb = \br{\Z{b}}/\br{\Z{q}}$ & $0.2170 \pm 0.0009$ &  \cite{LEPEWWG} \\
$\Rc = \br{\Z{c}}/\br{\Z{q}}$ & $0.1733 \pm 0.0048$ & \cite{LEPEWWG}\\
$\fs = \BR{\anti{b}{}{}}{\Bs}$ & $ 0.103^{+0.016}_{-0.015}$ & \cite{Schneider} \\ 
$\fd = \fu = \BR{\anti{b}{}{}}{\Bdu}$ & $ 0.395^{+0.016}_{-0.020}$ &\cite{Schneider} \\ 
$\fbb =  \BR{\particle{b}{}{}}{\mbox{\particle{b}{}{}-baryon}}$  & $ 0.106^{+0.037}_{-0.027}$ & \cite{Schneider} 
 \\[1pt] 
\hline 
$\BR{\particle{b}{}{}}{\ell}$                &  $0.1112 \pm 0.0020$ & \cite{LEPEWWG}\\
$\br{\particle{b}{}{} \rightarrow \particle{c}{}{} \rightarrow \ell}$ &  $0.0803 \pm 0.0033$ & \cite{LEPEWWG}\\
$\br{\particle{b}{}{} \rightarrow \anti{c}{}{} \rightarrow \ell}$ &  $0.0013 \pm0.0005$ & \cite{LEPHF}\\
$\BR{\particle{c}{}{}}{\ell}$                &  $0.098 \pm 0.005$ & \cite{LEPHF} \\
\hline
$\langle X_E \rangle$ & $0.702 \pm 0.008$ & \cite{LEPHF}  \\
\hline
\end{tabular}
\end{center}
\labt{phyparams}
\end{table}

In order to improve the rejection of non-\particle{b}{}{} background
or \particle{b}{}{} events with a badly estimated decay length error, 
the following additional cuts are applied~\cite{OL_thesis}:

\begin{itemize}

\item the momentum of the charm particle must be larger than 4~\GeVc; this cut is 
 increased to 8~\GeVc\ when the angle between the charm particle and the lepton is
less than $10^\circ$; 

\item the reconstructed mass of the \particle{b}{}{}-hadron must be less than 8~\GeVcc;  

\item the missing energy in the lepton hemisphere must be larger than $-2$~GeV;  

\item the angle between the charm particle and the lepton must be between $5^\circ$ and 
$30^\circ$;  

\item the angle between the charm particle and the jet must be less than $20^\circ$.
\end {itemize} 

Although the total efficiency of these additional requirements is 35\%, 
the average decay length resolution of the remaining events is 
improved by a factor of 2 and the amount of non-\particle{b}{}{} background 
in the sample reduced by a factor close to 4. In addition the average momentum 
resolution of the sample is significantly improved. 
A total of 33023 events survive after all cuts.  

\begin{table}
\tabcaption{Lepton candidate sources (\%), as estimated 
from Monte Carlo. Quoted uncertainties are statistical only.}
\labt{compo}
\begin{center}
\begin{tabular}{|c|c|c|c|c|}
\hline
 \Bs\ & \Bd\ & other \particle{b}{}{}-hadrons &  charm & \uds  \\
\hline
 $10.35 \pm 0.08$ & $38.53 \pm 0.13$ & $47.86 \pm 0.14$  & $2.31 \pm 0.06$ & 
$0.95\pm0.05$ \\
\hline
\end{tabular}
\end{center}
\end{table}

\begin{table}
\tabcaption{Definition of the eleven \Bs\ purity classes. 
Column 1 gives the number of charged tracks at the charm vertex. 
Column 2 shows whether the charge of these tracks 
are the same ({\it S}) or opposite ({\it O}) to that of the 
lepton, the tracks being ranked in order of decreasing momentum. 
Column 3 indicates the subclasses based on the presence of kaon or $\phi$
candidates at the charm vertex. 
Column 4 shows the fraction of data events in each class.
Column 5 gives the \Bs\ purity in each class, as estimated 
from Monte Carlo. Quoted uncertainties are statistical only.} 
\labt{enrichment}
\begin{center}
\begin{tabular}{|c|c|c|r@{\hspace{4.5ex}}|r@{\hspace{2ex}}|}
\hline
\begin{tabular}{@{}c@{}} Number   \\[-2pt] of tracks    \end{tabular} & 
\begin{tabular}{@{}c@{}} Charge   \\[-2pt] correlation  \end{tabular} & 
\begin{tabular}{@{}c@{}} Kaon   \\[-2pt] requirements \end{tabular} & 
\begin{tabular}{@{}c@{\hspace{-3.5ex}}} Fraction in \\[-2pt] data (\%) \end{tabular} & 
\multicolumn{1}{|c|}{\Bs\ purity (\%)} \\[6pt]
\hline\hline
   1   & $O$  & \multicolumn{1}{|@{}c@{}|}{\begin{tabular}{c}
                 1 kaon  \\ 0 kaon                            \end{tabular}}
              & \multicolumn{1}{|@{}r@{\hspace{4.5ex}}|}{\begin{tabular}{r@{}}
                 3.8  \\ 14.9               \end{tabular}}
              & \multicolumn{1}{|@{}r@{}|}{\begin{tabular}{r@{\hspace{2ex}}}
                24.0 $\pm$ 0.6 \\ 14.7 $\pm$ 0.3              \end{tabular}} \\
\hline
       & $OS,SO$  &   $\phi$    &  1.2 & 21.1 $\pm$ 1.0 \\
       & $OS,SO$  &   0 kaon    & 17.8 & 7.0  $\pm$ 0.2 \\
   2   & $OS,SO$  &   1 kaon    & 17.4 & 5.2  $\pm$ 0.1 \\
       & $OS,SO$  &  ~2 kaons   &  2.3 & 8.4  $\pm$ 0.5 \\
       \cline{2-5}
       & $OO$     &             &  8.3 & 16.7 $\pm$ 0.4 \\
\hline
       & $OOS$    &             &  2.9 & 19.4 $\pm$ 0.6 \\
   3   & $OSO$    &             &  3.8 & 18.0 $\pm$ 0.5 \\
       & $SOO$    &             &  3.9 & 14.5 $\pm$ 0.5 \\
\hline
\multicolumn{3}{|c|}{remainder} &  23.6 & 5.7 $\pm$ 0.1 \\
\hline

\end{tabular}
\end{center}
\end{table}

\section{\boldmath \Bs\ purity classes}\labs{enrichment}

\Table{compo} shows the composition of the final event sample obtained 
assuming the physics parameters listed in \Table{phyparams} 
and reconstruction efficiencies determined from Monte Carlo.
The average \Bs\ purity in the sample is estimated to be 10.35\%.

The sensitivity of the analysis 
to \Bs\ mixing is increased by splitting the data into subsamples 
with a \Bs\ purity larger or smaller than the average
and then making use of this knowledge in the likelihood fit.
Classes are constructed based on (i) the track 
multiplicity at the charm vertex, (ii) the number of identified kaon candidates 
and (iii) the charge correlation between the tracks at the charm vertex 
and the lepton.
The definition of the eleven classes used in this analysis
is given in \Table{enrichment}. 
As the last class contains those events which do not satisfy the 
criteria of the preceding classes,  
the classification procedure does not reject events.  
   
For an odd (even) number of charged tracks assigned to the charm vertex, 
the reconstructed charge of the b-hadron vertex is more likely to be zero
(non-zero), and therefore the probability for the hemisphere to contain
a neutral b-hadron is enhanced (reduced).
For events having two oppositely charged tracks at the charm vertex, the \Bs\ purity is 
6.7\%, which is lower than the average purity.
For this large subsample of events, the presence of kaon candidates
and consistency with the $\phi$ mass are used to recover some
sensitivity to the \Bs. 
In this procedure, kaon candidates are defined as charged tracks with
momentum above 2~\GeVc\ satisfying $\chipi+\chiK<0$ and $|\chiK|<2$,
and a $\phi$ candidate is defined as a pair of oppositely charged tracks
with an invariant mass between 1.01 and 1.03~\GeVcc\ (assuming kaon masses 
for the two tracks).

Monte Carlo studies indicate that this classification
procedure is effectively equivalent to increasing the statistics of the 
sample by  $28\%$.

\section{Proper time reconstruction and resolution}\labs{proper_time}

\begin{figure}
\vspace{-1.0cm}
\begin{center}
\makebox[0cm]{\psfig{file=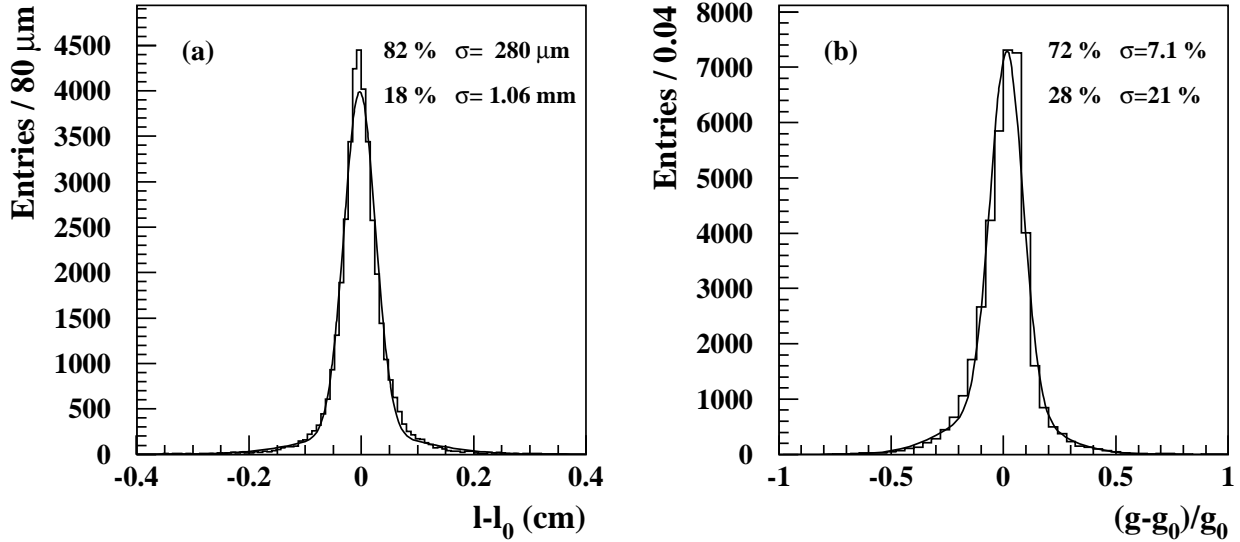,width=1.05\textwidth,
bbllx=0pt,bblly=271pt,bburx=560pt,bbury=560pt}}
\end{center}
\vspace{-1.0cm}
\figcaption{Decay length resolution (a) and relative boost term resolution (b)
for all \particle{b}{}{}-hadrons; 
the curves are the result of fits of the sum of two Gaussian functions
with relative fractions and widths as indicated.}
\labf{resol}
\end{figure}

\begin{table}[t]
\tabcaption{Double-Gaussian parametrizations of the decay length 
pull and relative boost term resolution obtained from Monte Carlo.}
\labt{lres_gres}
\begin{center}
\begin{tabular}{|c|c|c|}
\hline
\multicolumn{3}{|c|}{Parametrization of $(l-\lO)/\sigl$} \\
  $\alpha$ & Fraction $f_l^\alpha$  &  Sigma $\Sl^\alpha$  \\
\hline \hline
     1  &   $0.849 \pm 0.003$   & $ 1.333 \pm 0.005$  \\
     2  &   $0.151 \pm 0.003$   & $ 4.365 \pm 0.033$  \\
\hline
\end{tabular}
\hspace{0.5cm}
\begin{tabular}{|c|c|c|}
\hline
\multicolumn{3}{|c|}{Parametrization of $(g-\gO)/\gO$} \\
 $\beta$ &  Fraction $f_g^\beta$ &  Sigma $S_g^\beta$  \\
\hline \hline
  1 &  $0.723  \pm 0.004$  & $ 0.0713 \pm 0.0003$  \\
  2 &  $0.277  \pm 0.004$  & $ 0.2101 \pm 0.0012$  \\
\hline
\end{tabular}
\end{center}
\end{table}

An estimate, $l$, of the decay length of each \particle{b}{}{}-hadron candidate
is calculated as the distance from the 
interaction point to the \particle{b}{}{}-hadron vertex projected onto the direction 
of the jet associated to the lepton. 
This decay length includes a global correction of $-78~\mu$m,
determined using Monte Carlo events. This small offset is due to the
vertex reconstruction algorithm, which assumes that all lepton candidates
in \particle{b}{}{} events come from direct $\particle{b}{}{} \to \ell$ decays.
\Figure{resol}a shows the Monte Carlo distribution of $l-\lO$
for \particle{b}{}{} events,
where $\lO$ is the true decay length. 
An event-by-event decay length uncertainty, $\sigl$, is  
estimated from the covariance matrices of the tracks
attached to the vertices. This
can be compared with the true error, $(l-\lO)$, by
constructing the pull distribution, $(l-\lO)/\sigl$.
A fit to this Monte Carlo distribution of the sum of two Gaussian
functions  ($\alpha=1,2$) yields the
fractions, $f_l^\alpha$, and sigmas, $\Sl^\alpha$, indicated in \Table{lres_gres}.
These parameters are used to describe the observed tails when constructing
the resolution function.

The true boost term is defined as $\gO=\tO/\lO$,
where $\tO$ is the true proper time.
An estimate of the boost term is formed using
$g=m_{\mathrm B}/p_{\mathrm B} + 0.36~$ps/cm.
The average \particle{b}{}{}-hadron mass, $m_{\mathrm B}$, is assumed to be 5.3~\GeVcc\
and the reconstructed momentum is calculated as 
$p_{\mathrm B}=\sqrt{(E_{\mathrm c}+E_\nu+E_\ell)^2-m_{\mathrm B}^2}$
where $E_\ell$ is the measured lepton energy.
The constant term
is an average offset correction determined using Monte Carlo events; 
this results from the choice of the mass cut-off used in the nucleated
jet algorithm described in \Sec{eventsel}, which optimizes the relative boost
term resolution.
The distribution of $(g-\gO)/\gO$, shown in \Fig{resol}b,
is parametrized with the sum of
two Gaussian functions; \Table{lres_gres} shows the corresponding 
fractions, $f_g^\beta$, and sigmas, $S_g^\beta$, determined with 
Monte Carlo events.

\begin{figure}[tp]
\vspace{-2cm}           
\begin{center}
\makebox[0cm]{\psfig{file=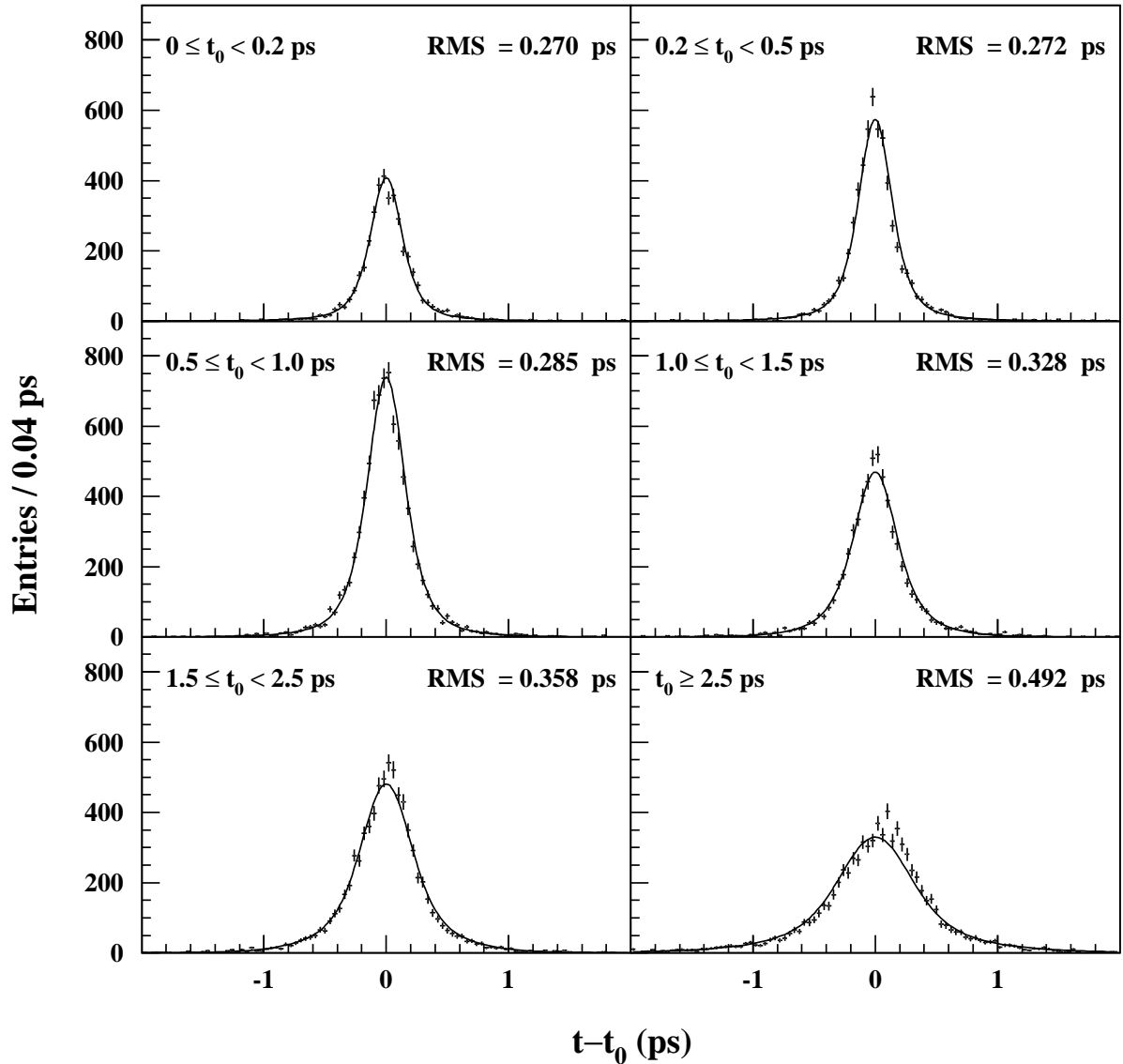,width=1.05\textwidth}}
\end{center}
\vspace{-1.0cm}
\figcaption{The proper time resolution for \particle{b}{}{} events in various intervals 
of true proper time $t_0$ (in ps). The curves display the corresponding
resolution assumed in the likelihood as obtained from \Eq{eqres}.
 The RMS values indicated, are derived from the data points shown.}

\labf{tres}
\end{figure}

The proper time of each \particle{b}{}{}-hadron candidate
is computed from the estimated decay length and boost term as
\begin{equation}
t = l g \, ,
\end{equation}
and its proper time resolution function is parametrized with 
the sum of four Gaussian components,
\begin{equation}
\Res(t,\tO) = \sum_{\alpha=1}^{2} \sum_{\beta=1}^2 f_l^{\alpha'} f_g^\beta
\frac{1}{\sqrt{2\pi} \sigma^{{\ab}} (\tO ) } 
\exp \left [ -\frac{1}{2} \left ( \frac{t-\tO}{\sigma^{\ab}(\tO )}
\right )^2
 \right ] \, , 
\labe{eqres}
\end{equation}
where $f_l^{2'}=\fldat f_l^2$ and $f_l^{1'}=1-f_l^{2'}$, and 
where the event-by-event resolution $\sigma^{\alpha\beta}$ 
of each component, given by
\begin{equation}
\sigma^{\ab} (\tO) = \sqrt{  
  \left(g \Sldat \Sl^\alpha \sigl\right)^2 
+ \left(\tO \Sgdat S_g^\beta \right)^2   } \, ,
\labe{treso}
\end{equation}
includes the explicit dependence on $\tO$.
This parametrization 
implicitly assumes that any correlation between the decay length resolution
and the relative boost resolution is small, as confirmed by Monte Carlo
studies.

The scale factors $\Sldat$ and $\fldat$ 
are introduced to account for a possible discrepancy between data and
Monte Carlo, both in the amount of tail in the decay length pull
($\fldat$) and in the estimate of $\sigl$ itself ($\Sldat$).
In a similar fashion, the inclusion 
of the parameter $\Sgdat$
allows possible systematic uncertainties due to the boost resolution
to be studied.
By definition all
these factors are set to unity when describing the resolution of simulated
events.
\Figure{tres} shows, for various intervals of true proper time, the proper
time resolution of simulated \particle{b}{}{} events together with
the parametrization obtained from \Eq{eqres}.
The parametrization is satisfactory, especially for small proper times.

\begin{figure}[t]
\begin{center}
\vspace{-0.5cm}
\psfig{file=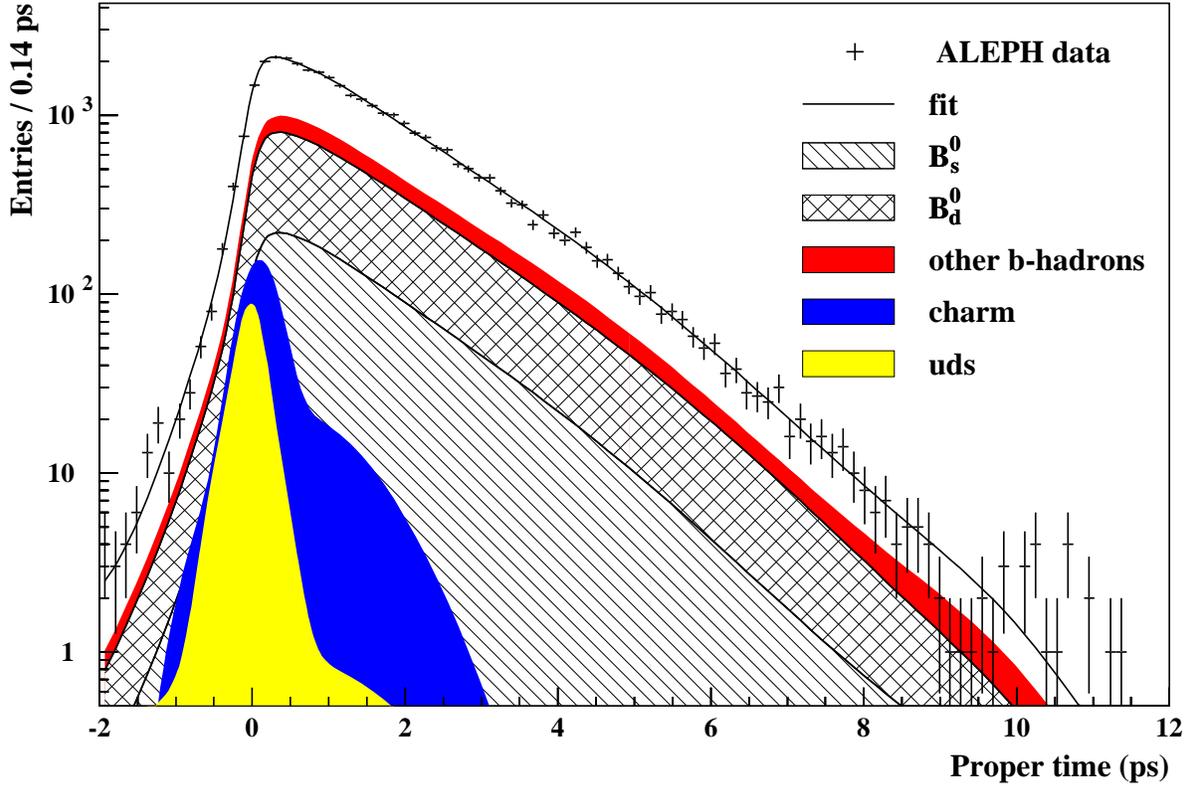,width=1.\textwidth,
bbllx=0pt,bblly=189pt,bburx=560pt,bbury=560pt}
\end{center}
\vspace{-0.7cm}
\figcaption{The reconstructed proper time distributions of the selected 
events in data. 
The contributions from the various components are indicated. The curve is
the result of the fit described in \Sec{proper_time}.}
\labf{fig_lifetime}
\end{figure}

In order to measure $\Sldat$ and $\fldat$ in the data, 
a fit is performed to the reconstructed proper time 
distribution of the selected sample of real events.
This is performed using
the likelihood function described in \Sec{likelihood}, modified 
to ignore tagging information.
Fixing all physics parameters to their central values given
in \Table{phyparams},
the likelihood is maximized with respect to $\Sldat$ and $\fldat$.
The fit reproduces well the 
negative tail of the proper time distribution (see \Fig{fig_lifetime}),
showing that the resolution is satisfactorily described by the chosen
parametrization.
The fitted values $\Sldat=1.02\pm0.03$ and $\fldat=1.20\pm0.09$
indicate that the decay length resolution in the data is somewhat worse than 
that suggested by the Monte Carlo simulation.

\section{Initial and final state tagging} \labs{tagging}

\def\Qoppo{Q_{\mathrm o}}
\def\Qsame{\tilde{Q}_{\mathrm s}}
\def\ptl{${p_{\rm T}^\ell}$}
\def\sq{$S(\Qoppo)$}
\def\sk{$S(\K)$}
\def\sl{$S(\ell)$}
\def\slo{$S(\ell_{\mathrm o})$}
\def\sls{$S(\ell_{\mathrm s})$}
\def\sqoqs{${\Qsame\times S(\Qoppo)}$}
\def\sqoqo{${\Qoppo\times S(\Qoppo)}$}
\def\skqs{${\Qsame\times S(\K)}$}
\def\skqo{${\Qoppo\times S(\K)}$}
\def\zk{${Z_{\k}}$}
\def\slqo{${\Qoppo\times S(\ell)}$}
\def\slqs{${\Qsame\times S(\ell)}$}
\def\xeff{x^{\mbox{\scriptsize eff}}}

The flavour state of the decaying \Bs\ candidate
is estimated from the charge of the reconstructed lepton.
This final state tag is incorrect if the
lepton is from the $\particle{b}{}{} \rightarrow \particle{c}{}{} 
\rightarrow \ell$ decay
(6.1\% of the \particle{b}{}{} events in the sample) as the charge of the 
lepton is reversed.
The flavour state at production is estimated using three initial state tags.
A \Bs\ candidate is ``tagged as unmixed (mixed)'' when the
reconstructed initial and final flavour states are the
same (different).
By definition, candidates from charm, \uds,
or non-oscillating \particle{b}{}{}-hadron backgrounds are correctly tagged
if they are tagged as unmixed.

The tagging power
is enhanced by the means of discriminating variables which
have some ability to distinguish between correctly tagged and 
incorrectly tagged candidates.
This approach was first used in the ALEPH 
\Ds--lepton analysis~\cite{ALEPH-DS-LEPTON} and refined for the 
\Ds--hadron analysis~\cite{ALEPH-DSHAD}.
In contrast to what was performed in \Refss{ALEPH-DS-LEPTON}{ALEPH-DSHAD},
an event is considered to be mistagged if either the initial or final state is 
incorrectly tagged, but not both.

\begin{figure}
\begin{center}
\mbox{\psfig{file=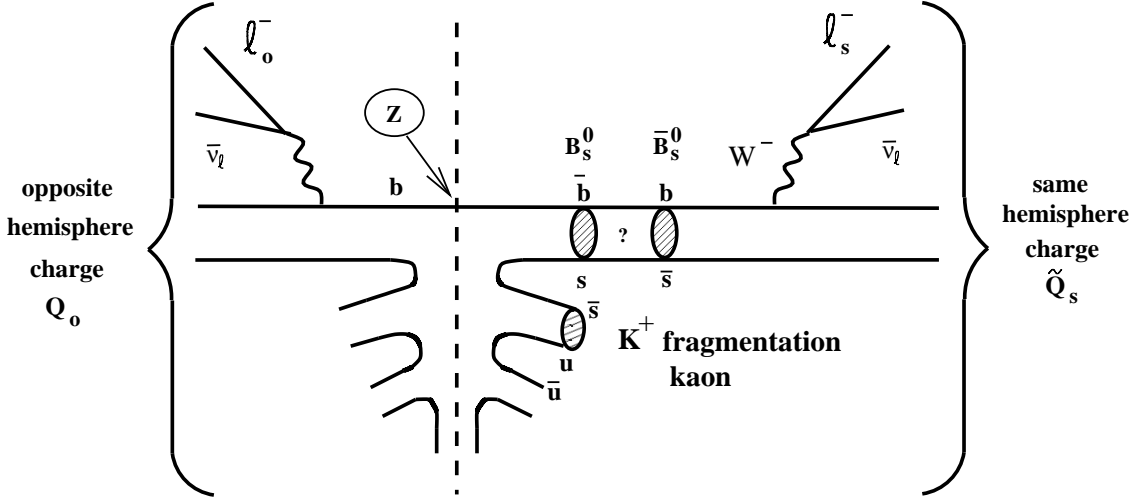,width=15cm}}
\end{center}
\figcaption{Schematic drawing indicating the initial and final state 
tags used in this analysis.}
\labf{fig_decay}
\end{figure}

For each \Bs\ candidate, one of the tags described below is 
used to determine the initial state (see also \Fig{fig_decay}).

\begin{itemize}
\item {\bf Opposite lepton tag:}
Leptons with momentum larger than 3~\GeVc\ are searched for in the 
hemisphere opposite to the \Bs\ candidate. 
The sign of the lepton with the highest transverse momentum $p_T(\ell_{\mathrm o})$  
tags the nature of the initial b quark in the opposite hemisphere. 
It takes precedence over the other tags if it is available.

\item {\bf Fragmentation kaon tag:}
The fragmentation kaon candidate is defined as
the highest momentum charged track within $45^{\circ}$ of
the \Bs\ direction, identified, using the vertexing algorithm described 
in Section 2, as being more likely to come 
from the interaction point than the charm vertex, and
satisfying $\chiK < 0.5 $ and $\chiK - \chipi > 0.5$.

The sign of the fragmentation kaon candidate tags the sign of the b quark
in the same hemisphere. It is used if no opposite 
hemisphere lepton tag is found.

\item {\bf Opposite hemisphere charge tag:}
The opposite hemisphere charge is defined as
\begin{equation}
\Qoppo =
\frac{\displaystyle \sum_i^{\rm oppo} q_i \, \vert p^i_{\parallel} \vert ^{\kappa}}
     {\displaystyle \sum_i^{\rm oppo}     \vert p^i_{\parallel} \vert ^{\kappa}} \, ,
\labe{Qoppo}
\end{equation}
where the sum is over all charged particles in the opposite hemisphere,
$p^i_{\parallel}$ is the momentum of the $i^{\rm th}$
track projected on the thrust axis, $q_i$ its charge and
$\kappa = 0.5$. The sign of $\Qoppo$
tags the initial state of the b quark in the opposite hemisphere.
This tag is always available but has the largest mistag probability of
the three tags. It is used only if no other tag is available.
\end{itemize}

\newcommand{\YES}{used}
\newcommand{\no}{}
\begin{table}
\begin{center}
\figcaption{The tag and discriminating variables used in each class.
The quantities \sq, \sk\ and \slo\ are the signs of the 
opposite hemisphere charge, the fragmentation kaon and the opposite
side lepton. Classes 3--5 all use the sign of the opposite hemisphere
lepton as the initial state tag. For Class 3 no fragmentation kaon candidate 
is identified. For Class 4 (Class 5) a fragmentation kaon candidate is found 
whose charge is the same as (opposite to) the charge of the opposite 
hemisphere lepton.
Purity and mistag rates are estimated from Monte Carlo.
Quoted uncertainties are statistical only.}
\labt{classes}
\vspace{-0.5cm}
\begin{tabular}{|c|c|c|c|c|c|}
\hline
Tagging class & 1 & 2 & 3 & 4 & 5  \\ 
\hline
Available initial & \sq\ & \sq\ & \sq\  & \sq\ & \sq\  \\
state tags        &    & \sk\ & \slo\ & \slo=--\sk\ & \slo=\sk\   \\
\hline\hline
Intial state tag used & \sq\ & \sk\ & \slo\ & \slo=--\sk\ & \slo=\sk\   \\
\hline

              & $|\Qoppo|$     & \no        & \no         & \no         & \no         \\
              & $S(\Qoppo)\Qsame$ & \no        & \no         & \no         & \no         \\
              & \no         & $S(K)\Qoppo$  & \no         & \no         & \no         \\
              & \no         & $S(K)\Qsame$  & \no         & \no         & \no         \\
Discriminating& \no         & $\chi_\pi$ & \no          & $\chi_\pi$  & $\chi_\pi$  \\
variables     & \no         & \zk\       & \no          &  \zk\       & \zk\       \\
used          & \no         & \no        & \slo$\Qoppo$ & \slo$\Qoppo$ & \slo$\Qoppo$ \\
              & \no         & \no        & \slo$\Qsame$ & \slo$\Qsame$ & \slo$\Qsame$ \\
              & \no         & \no        & $p_T(\ell_{\mathrm o})$  & $p_T(\ell_{\mathrm o})$  & $p_T(\ell_{\mathrm o})$  \\
              & \no         & $t$        & \no         & $t$         & $t$         \\
              & $p_T(\ell_{\mathrm s})$  & $p_T(\ell_{\mathrm s})$ & $p_T(\ell_{\mathrm s})$  & $p_T(\ell_{\mathrm s})$  & $p_T(\ell_{\mathrm s})$  \\

\hline \hline
Fraction in data (\%)& 71.4 $\pm$ 0.2 & 11.9 $\pm$ 0.2 & 14.2 $\pm$ 0.2 & 1.3
$\pm$ 0.1 & 1.2 $\pm$ 0.1  \\

\hline
 \Bs\ purity (\%)& 9.8 $\pm$ 0.1 & 13.1 $\pm$ 0.3 & 10.1 $\pm$ 0.2 &
 15.6 $\pm$ 1.0 &  11.8 $\pm$ 0.8 \\

 \Bs\ mistag (\%)& 38.6 $\pm$ 0.5 & 28.9 $\pm$ 1.0 & 34.0 $\pm$ 1.1 & 
16.1 $\pm$ 2.3 & 55.9 $\pm$ 3.5  \\

$\!\!\!$ \Bs\ effective mistag (\%) $\!\!\!$& 32.4 & 24.0 & 24.5 & 12.5 & 22.3  \\

\hline
 \Bd\ mistag (\%)& 38.4 $\pm$ 0.2 & 48.5 $\pm$ 0.7 & 35.4 $\pm$ 0.5 & 
35.5 $\pm$ 2.0 & 39.9 $\pm$ 2.0  \\

 other B mistag (\%)& 37.6 $\pm$ 0.2 & 61.4 $\pm$ 0.5 & 34.2 $\pm$ 0.5 & 
43.8 $\pm$ 1.8 & 24.1 $\pm$ 1.4  \\

 charm mistag (\%)& 38.2 $\pm$ 1.4 & 54.2 $\pm$ 3.2 & 14.2 $\pm$ 3.1 & 
 50.0 $\pm$ 50.0 & 8.6 $\pm$ 8.2  \\

 \uds\ mistag (\%)& 47.8 $\pm$ 2.8 & 56.9 $\pm$ 6.0 & 46.0 $\pm$ 12.9 & 
 50.0 $\pm$ 50.0 & 50.0 $\pm$ 50.0  \\

\hline
\end{tabular}
\end{center}
\end{table}

The events are sorted into five exclusive classes
based on the availability and results of the three tags.
The definition of these tagging classes and the list of the discriminating
variables associated with each class are given in \Table{classes}. 
The variable $\Qsame$
is the sum of the charges of all the tracks in the same hemisphere
and carries information on the initial state of the \Bs.
As the sum of charges of tracks originating from the decay of a neutral
particle is zero,
it is independent of whether the \Bs\ decays as a \Bs\ or a \Bsbar.
The variable \zk\ is defined as \zk$= p_{\mathrm K} / (E_{\mathrm beam} - E_{\mathrm B})$,
 where $p_{\mathrm K}$ is the kaon momentum,
$E_{\mathrm beam}$ the beam energy and $E_{\mathrm B}$ the \Bs\ candidate energy.
The inclusion of the reconstructed \Bs\ proper time $t$
takes into account that the mistag probability of the fragmentation kaon tag
increases as the \Bs\ vertex approaches the primary vertex, 
due to the misassignment of tracks between the primary and 
secondary vertices. The use, for all classes, of the variable $p_T(\ell_{\mathrm s})$,
the transverse momentum of the lepton from the \Bs\ candidate decay,
reduces the deleterious effect of 
$\particle{b}{}{} \rightarrow \particle{c}{}{} \rightarrow \ell$ on the final
state mistag.

The mistag probability, $\eta$, for the \Bs\ signal events in each class,
as well as the probability distributions
of each discriminating variable $x_i$
for correctly and incorrectly tagged signal events,
$r_i(x_i)$ and $w_i(x_i)$,
are estimated from Monte Carlo. 
The various discriminating variables chosen in each class, $x_1, x_2, \ldots$,
are combined into a single effective discriminating variable $\xeff$, 
according to the prescription developed for the
\Ds\ based analyses~\cite{ALEPH-DS-LEPTON,ALEPH-DSHAD}.
This new variable is defined as
\begin{equation}
\xeff = \frac{
   \eta  \, w_1(x_1) \, w_2(x_2) \, \cdots }{
(1-\eta) \, r_1(x_1) \, r_2(x_2) \, \cdots \, +
   \eta  \, w_1(x_1) \, w_2(x_2) \, \cdots } \, ,
\labe{xeff}
\end{equation}
and takes values between 0 and 1.
A small value indicates that the \Bs\ oscillation is 
likely to have been correctly tagged.

To allow use of the discriminating variables in the likelihood fit, 
the probability density functions $G^c_{jkl}(\xeff)$ of $\xeff$ are
determined for each lepton source $j$, in each tagging class $k$ and in each 
\Bs\ purity class $l$,
separately for the correctly ($c=+1$) and incorrectly ($c=-1$) tagged events.
This determination (as well as the estimation of the corresponding 
mistag probabilities $\eta_{jkl}$) is based on Monte Carlo.

The enhancement of the tagging power provided by the variable, $\xeff$, 
depends on the difference between the $G^+_{jkl}(\xeff)$ and $G^-_{jkl}(\xeff)$
distributions, and can be quantified in terms of
effective mistag rates, as described in \Ref{ALEPH-DS-LEPTON}. 
The effective mistag rates for the \Bs\ signal in the five tagging classes 
are given in \Table{classes}. 
This table also indicates \Bs\ purity and the mistags for 
all background components. 
The overall average \Bs\ effective mistag is 29\%.

\Figure{fig_xeffcomp}
displays the distribution of $\xeff$ in each of the tagging classes;
a good agreement is observed between data and Monte Carlo.
The systematic uncertainties associated with the 
tagging procedure are considered in \Sec{sec_syst}.

\begin{figure}
\begin{center}
\vspace{-2.0cm}
\mbox{\psfig{file=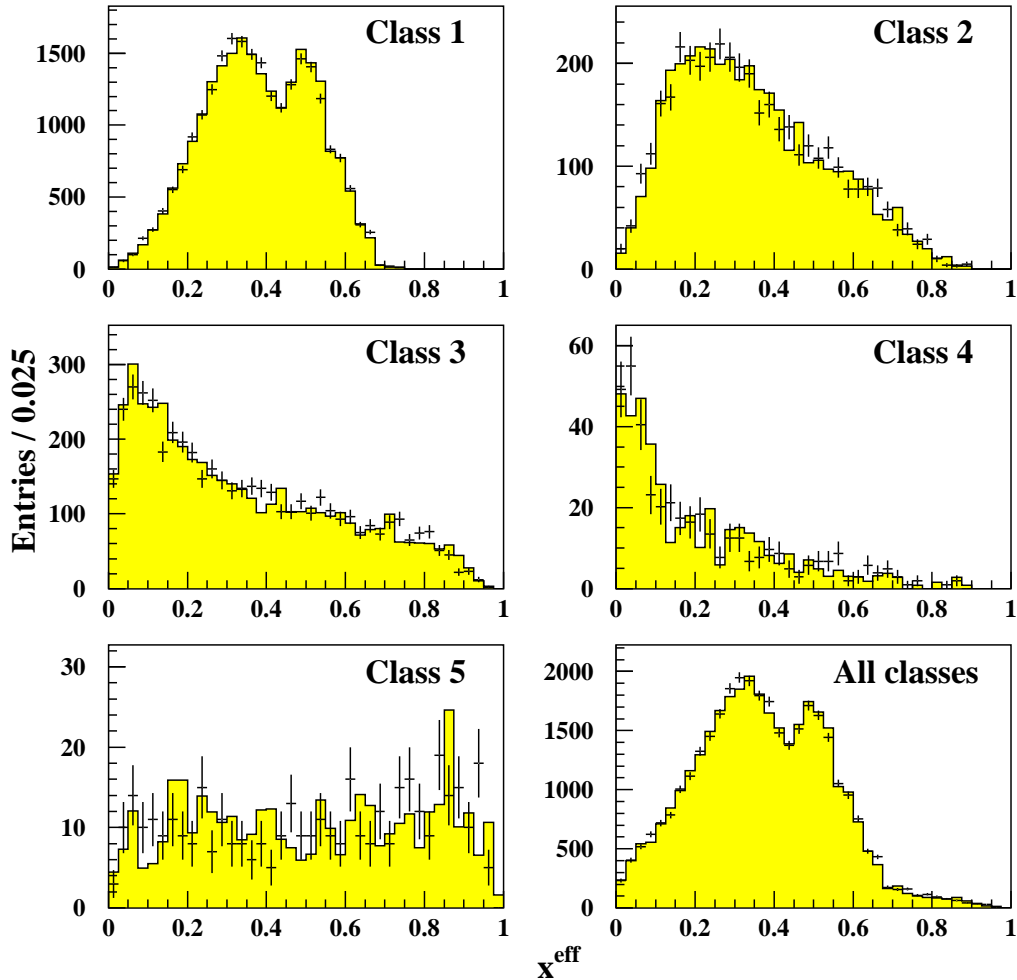,width=15cm}}
\end{center}
\figcaption{Distribution of $\xeff$ in each
tagging class in data (points) and Monte Carlo (histogram).}
\labf{fig_xeffcomp}
\end{figure}

\section{Likelihood function} \labs{likelihood}

Each \particle{b}{}{}-hadron source has a different probability distribution function
for the true proper time, $t_0$, and for
the discrete variable, $\muO$, defined to take the value $-1$
for the mixed case or $+1$ for the unmixed case.
Assuming CP conservation and equal decay widths
for the two CP eigenstates in each neutral \particle{B}{}{}-meson system, the joint 
probability distribution of $t_0$ and $\muO$ 
can be written as
\begin{equation}
p_j(\muO,t_0) =
     \frac{e^{-t_0/\tau_j}}{2 \tau_j} \,
     \left[1 + \muO \cos{(\Delta m_j \, t_0)}
     \right] \, ,
     \labe{p}
\end{equation}
where $\tau_j$ and $\Delta m_j$ are the lifetime and oscillation frequency of
\particle{b}{}{}-hadron source $j$
(with the convention that $\Delta m_j = 0$ for non-oscillating \particle{b}{}{}-hadrons).

The efficiency for reconstructing the 
\particle{b}{}{}-hadron vertex depends on the true proper time.
The stringent selection cuts described in \Sec{eventsel} 
are designed to reduce the fraction 
of fragmentation tracks assigned to the charm vertex,  
consequently causing a loss of efficiency at small proper times. 
Similarly at large proper times the efficiency also decreases as one is 
less likely to include a fragmentation track at the charm vertex 
and therefore more likely to fail the requirement of the charm vertex  
being assigned at least one track. 
The efficiencies $\eps_j (\tO)$ are parametrized separately
for each \particle{b}{}{}-hadron component $j$. They are
independent of whether the \particle{b}{}{}-hadron candidate is tagged as mixed or unmixed. 

The joint probability distribution of the reconstructed
proper time $t$ and of $\muO$ is obtained as the convolution of
$p_j(\muO,t_0)$ with the event-by-event resolution function $\Res(t,\tO)$ 
(\Sec{proper_time}) and takes into account the observed dependence of 
the selection efficiency on true proper time:
\begin{equation}
h_{j}(\muO,t)  = \frac{\displaystyle
     \int_0^{\infty} \eps_j (\tO)
     p_j(\muO,\tO)   \Res (t,\tO)
     \,d\tO }
   {\displaystyle \int_0^{\infty} \eps_j (\tO)
      \frac{1}{\tau_j}e^{-\tO/\tau_j}
     \,d\tO } \, .
\labe{eqh}
\end{equation}

For the lighter quark backgrounds,
$h_{j}(-1,t)=0$ as these sources are unmixed by definition, and
$h_{j}(+1,t)$ are the reconstructed proper time distributions.
These distributions are determined from Monte Carlo samples
and are parametrized as the sum of three Gaussian functions.

The likelihood function used in this analysis is based on the values
taken by three
different variables in the selected data events.
These variables are the reconstructed proper time $t$,
the tagging result $\mu$, taking the value
$-1$ for events tagged as mixed or $+1$ for those tagged as unmixed,
and the effective discriminating variable $\xeff$.
The use of the discriminating variable
$\xeff$ in the likelihood function is reduced to the use of two sets of
functions of $\xeff$, $X_{jkl}(\xeff)$ and $Y_{jkl}(\xeff)$ (described below),
 whose values can be interpreted as event-by-event mistag probabilities and
fractions of the different lepton sources respectively.
The likelihood of the total sample is written as
\begin{equation}
\like = C \prod_l^{\scriptsize \mbox{~11 purity~}}
          \prod_k^{\scriptsize \mbox{~5 tagging~}}
          \prod_i^{\scriptsize \mbox{~$N_{kl}$ events~}}     
          f_{kl}(\xeff_{ikl},\mu_{ikl},t_{ikl}) \, ,
\labe{eqlike}
\end{equation}
where $C$ is a constant independent of \particle{b}{}{}-hadron oscillation
frequencies and lifetimes, $N_{kl}$ is the number of selected candidates
from \Bs\ purity class $l$ falling in tagging class $k$, and where
\begin{equation}
f_{kl}(\xeff,\mu,t) = \sum_j^{\scriptsize \mbox{5 sources}}
Y_{jkl}(\xeff) \left[
\left(1-X_{jkl}(\xeff)\right) h_{j}(\mu,t) + 
   X_{jkl}(\xeff) h_{j}(-\mu,t)  \right]
\,
\labe{pdf}
\end{equation}
sums over the 5 different lepton sources considered to comprise the sample (see \Table{compo}). 
The event-by-event quantities $X_{jkl}(\xeff)$ and $Y_{jkl}(\xeff)$
are computed from the 
distributions $G^c_{jkl}(\xeff)$ and mistag probabilities $\eta_{jkl}$
introduced in \Sec{tagging},
\begin{equation}
X_{jkl}(\xeff) = \eta_{jkl} \, \frac{G^-_{jkl}(\xeff)}{G_{jkl}(\xeff)} \, ,
\mbox{~~~}
Y_{jkl}(\xeff) = \alpha_{jkl} \, \frac{G_{jkl}(\xeff)}{\sum_{j'} 
\alpha_{j'kl} G_{j'kl}(\xeff)} \, ,
\end{equation}
where $G_{jkl}(\xeff) = (1-\eta_{jkl}) G^+_{jkl}(\xeff) + 
\eta_{jkl} G^-_{jkl}(\xeff)$
and where $\alpha_{jkl}$ are the source fractions,
satisfying $\sum_{j=1}^{\mbox{\scriptsize 5 sources}} \alpha_{jkl} = 1$.

\section{Results for \boldmath \dms} \labs{results}

\begin{figure}
\vspace{-2cm}
\begin{center}
\makebox[0cm]{\psfig{file=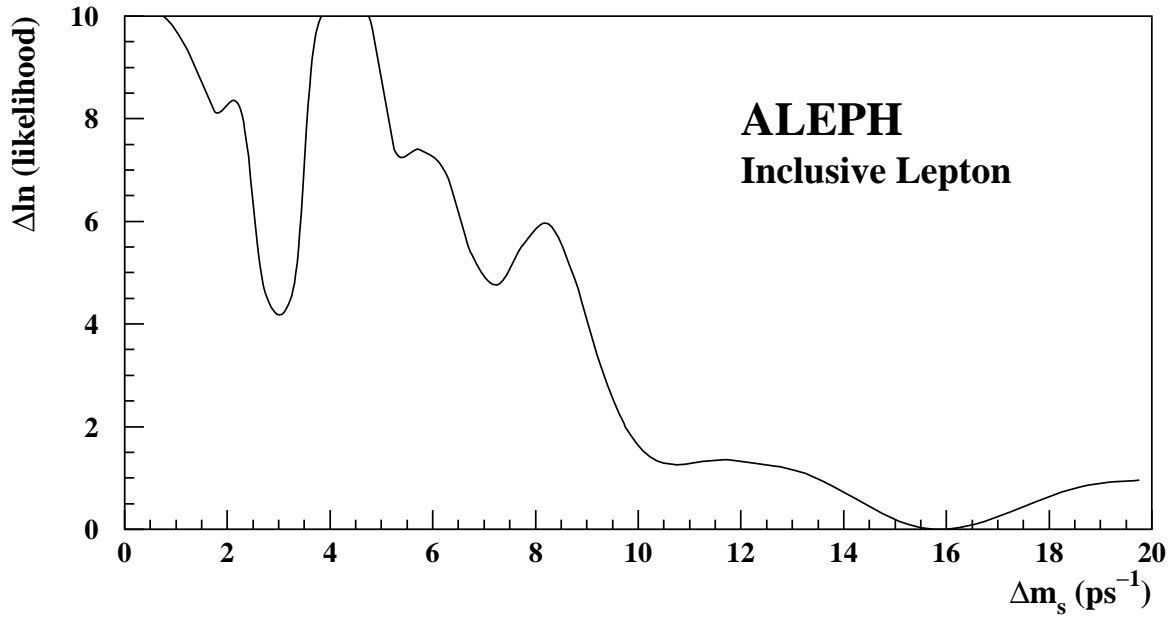,width=1.05\textwidth,
bbllx=0pt,bblly=266pt,bburx=560pt,bbury=560pt}}
\end{center}
\figcaption{Negative log-likelihood difference with respect to the minimum 
as a function of \dms.}
\labf{likel}
\end{figure}

\begin{figure}
\vspace{-2.3cm}
\begin{center}
\makebox[0cm]{\psfig{file=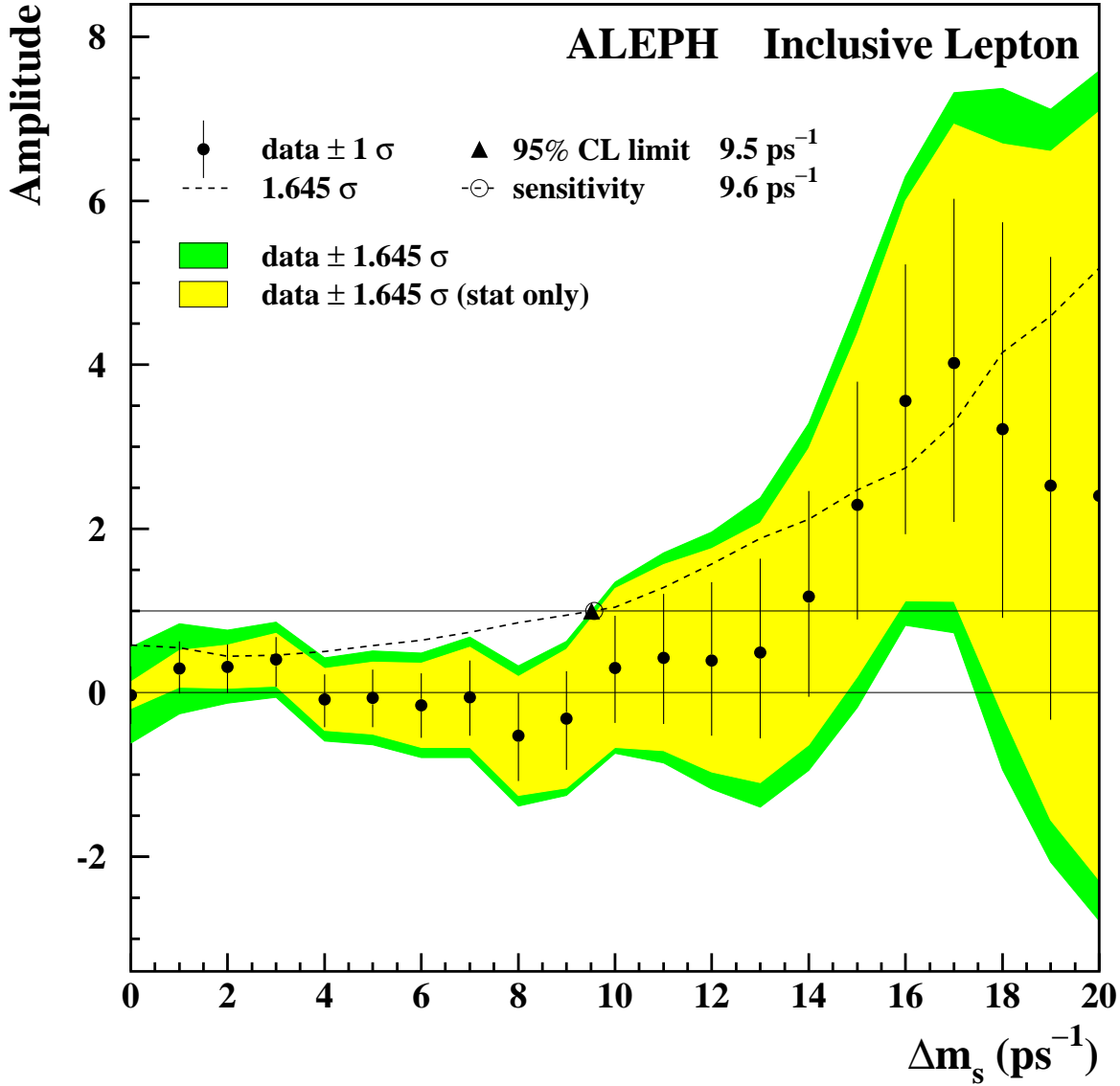,width=1.05\textwidth}}
\end{center}
\vspace{-0.6cm}
\figcaption{Measured \Bs\ oscillation amplitude as a function of \dms\ for this analysis.
The error bars represent the 1$\sigma$ total uncertainties, and
the shaded bands show the one-sided \CL{95} contour, with and without 
systematic effects included.}
\labf{amplitude}
\end{figure}

Assuming the values for the physics parameters given in \Table{phyparams},  
the variation in the data of the log-likelihood, 
as a function of the free parameter \dms, is shown in \Fig{likel}. 
The difference in log-likelihood is plotted relative to its global minimum and
remains constant for \dms\ larger than 20~\invps. 
The global minimum occurs at $\dms=15.9\pm1.6$(stat.)~\invps\ but is not 
sufficiently deep to claim a measurement.

In order to extract a lower limit on \dms\ and to facilitate combination with
other analyses, the results are also presented in terms of 
the ``amplitude'' fit. In this method~\cite{HGM_NIM} the
magnitude of \Bs\ oscillations is measured at fixed values
of the frequency \dms, using a modified likelihood function that 
depends on a new parameter, the \Bs\ oscillation amplitude ${\cal A}$.
This is achieved by replacing the probability density function of
the \Bs\ source given in \Eq{p} with
\begin{equation}
   \frac{e^{-t_0/\taus}}{2 \taus} \,
   \left[1 + \muO {\cal A} \cos{(\dms \, t_0)}
   \right] \, .
\end{equation}
For each value of \dms, the new negative log-likelihood is then minimized with
respect to ${\cal A}$, leaving all other parameters (including \dms) fixed.
The minimum is well behaved and very close to parabolic. At each value of 
\dms\ one can thus obtain a measurement of the amplitude with Gaussian error, 
${\cal A} \pm \sigAstat$.
If $\dms$ is close to the true value, one expects ${\cal A} = 1 $ within 
the estimated uncertainty; however, if \dms\ is far from its
true value, a measurement consistent with ${\cal A} = 0$ is expected.

The amplitude fit results are displayed in \Fig{amplitude} 
as a function of \dms. A peak in the amplitude, corresponding to the 
minimum observed in the negative log-likelihood, can be seen around 
$\dms=16~\invps$. At this value, the  measured amplitude
is $2.2\,\sigma$ away from zero; 
as for the likelihood, 
this is not significant enough to claim a measurement of \dms. 

A value of \dms\ can be excluded at \CL{95} if 
${\cal A}+1.645~\sigma_{\cal A}<1$.
Taking into account all systematic uncertainties described
in the next section, all values of \dms\ below 9.5~\invps\ are excluded
at \CL{95}. The  sensitivity, estimated from the data as the value
of \dms\ at which $1.645\,\sigma_{\cal A}=1$, is 9.6~\invps.
Ignoring systematic uncertainties would increase the \CL{95} lower limit 
and sensitivity by 0.1~\invps\ and 0.6~\invps\ respectively.

\section{Systematic uncertainties}
\labs{sec_syst}

The systematic uncertainties on the \Bs\ oscillation amplitude $\sigAsyst$
 are calculated, using the prescription of \Ref{HGM_NIM}, as 
\begin{equation}
\sigAsyst ={\cal A}^{\mbox{\scriptsize new}}-
{\cal A}^{\mbox{\scriptsize nom}}
 ~ + ~ (1-{\cal A}^{\mbox{\scriptsize nom}} )
\frac{\sigma^{\mbox{\scriptsize new}}_{{\cal A}}-
\sigma^{\mbox{\scriptsize nom}}_{{\cal A}}} 
{\sigma^{\mbox{\scriptsize nom}}_{{\cal A}} }
\end{equation}
where the superscript ``\mbox{\small nom}'' 
refers to the amplitude values and statistical 
uncertainties obtained
using the nominal values for the various parameters and 
``\mbox{\small new}'' refers 
to the new amplitude values obtained when a single parameter is changed and 
the analysis repeated (including a re-evaluation of the distributions of the 
discriminating variables used for the \particle{b}{}{}-flavour tagging). 
The total systematic uncertainty is the quadratic sum of the
following contributions.
  
\begin{itemize} 
\item{\bf Sample composition:}
The systematic uncertainty on the sample composition is obtained by 
varying the assumed values for the \particle{b}{}{}-hadron fractions \fs, $\fbb$ and 
the various lepton sources ($\particle{b}{}{} \rightarrow \ell$, 
$\particle{b}{}{} \rightarrow \particle{c}{}{} \rightarrow \ell$, etc \ldots) by 
the uncertainties quoted in \Table{phyparams}. 
The statistical uncertainty on the purities determined from  Monte Carlo 
is also propagated.

A comparison of data and Monte Carlo fractions for the different \Bs\ 
purity classes shows small deviations, the largest relative difference of 
16\%  occurring in the first class of  \Table{enrichment}.
The systematic uncertainty due to the \Bs\ purity classification procedure 
is evaluated by shifting, in each class, all five 
purities (\Bs, \Bd, \ldots)
in the direction of their respective overall averages,
$\overline{\alpha}_j$ given in \Table{compo}, by a fraction
$\gamma =\pm 20\%$ of their differences with respect to these averages:
\begin{equation}
\alpha_{jkl} \rightarrow \alpha_{jkl} + \gamma (\alpha_{jkl} - \overline{\alpha}_j) \, .
\end{equation} 
As this is performed coherently in 
all \Bs\ purity classes, the procedure is rather conservative and ensures 
that the overall average purities remain unchanged.
Not using the \Bs\ purity classification would decrease the \dms\ statistical 
sensitivity by 0.7~\invps.

For the fraction of charm and \uds\ backgrounds a relative variation of $\pm 25\%$ 
is propagated, as suggested from a comparison between data and Monte Carlo
performed in \Ref{high_pt_lepton}.

\item{\bf Proper time resolution:}
For the systematic uncertainty on the proper time resolution,
the correction factors presented in Tables 4 and 5 are varied by $\pm1\sigma$.
The scale factors 
($\Sldat=1.02\pm0.03$ and $\fldat=1.20\pm0.09$) for the decay length resolution,
obtained from the lifetime fit to the data, are also varied by their measured
uncertainty.
In addition, a possible bias of $\pm 0.055$~ps/cm is considered on the 
determination of the boost term; this value corresponds to the observed 
shift between the measured and simulated boost term distributions and 
represents approximately 1\% of the average boost term.
Finally the boost term resolution is given a relative variation of $\pm10\%$ 
($\Sgdat=1.0\pm0.1$), which is conservative given the close agreement between
the measured and simulated boost distributions.
 
\item{\bf \boldmath \particle{b}{}{}-quark fragmentation:}
The average fraction of energy taken by a \particle{b}{}{}-hadron during the 
fragmentation process,
$\langle X_E \rangle =0.702\pm0.008$,
is varied by its measured uncertainty. 
The corresponding effects on the 
sample composition, mistags and resolutions are propagated. 
 
\item{\bf Mistag:}
Based on data/Monte Carlo comparisons of the tagging variables, 
performed for the \Ds-based analyses~\cite{ALEPH-DS-LEPTON,ALEPH-DSHAD}, 
absolute variations of $\pm0.8\%$ for the first tagging class
(opposite hemisphere charge)
and $\pm2\%$ for all other classes (fragmentation kaon and opposite lepton) are applied to the mistag rates.
In addition, the $\pm 1\sigma$ statistical uncertainty from Monte Carlo is 
propagated. 

The changes in mistag due to variation 
of the $\particle{b}{}{} \rightarrow \particle{c}{}{} \rightarrow \ell$
fraction are included as part of the sample composition systematic uncertainty. 

\item{\bf Lifetimes, \boldmath{\dmd, \Rb\ and \Rc}:}
The values assumed for the various \particle{b}{}{}-hadron lifetimes, \dmd, \Rb\ and \Rc\
are varied within the uncertainties quoted in \Table{phyparams}. 

\item{\bf Difference in decay width:} 
A possible decay width difference \dggs\  between the two mass 
eigenstates of the \Bs\ meson has been ignored in the likelihood fit. The fit 
is therefore repeated with a modified likelihood assuming $\dggs = 0.27$,   
equal to the theoretical prediction of \Ref{DELTA_GAMMA},
$\dggs = 0.16^{+0.11}_{-0.09}$, plus its quoted positive uncertainty.

\item{\bf Cascade bias:}
In the likelihood expression of \Eq{eqlike} each \particle{b}{}{}-hadron component
is treated using a single resolution function and mistag. No attempt is 
made to treat separately the $\particle{b}{}{} \rightarrow \ell$ (direct)
and \mbox{$\particle{b}{}{} \rightarrow \particle{c}{}{} \rightarrow \ell$}
(cascade) decays. 
While the former is characterized by a good proper time resolution and mistag, 
the latter has a degraded decay length resolution and a somewhat biased
decay length because of the charm lifetime. In addition, the sign of the lepton
is changed, leading to a different total mistag.
To study the possible bias arising from the correlation between the poor decay length 
resolution and degraded tagging performance of the cascade events, 
two different fast Monte Carlo experiments are generated 
with a true value of \dms\ equal to $50~\invps$. In the first 
the \particle{b}{}{}-hadron decays are generated using the average mistag and resolution;
in the second, the primary and cascade components are generated separately, 
each with their appropriate mistag and resolution. 
For both experiments, the corresponding amplitude plot is obtained 
using the likelihood described in \Sec{likelihood}, i.e.\ with
average mistags and resolutions.

The fast Monte Carlo experiment 
generated using the average \particle{b}{}{}-hadron properties, 
yields an amplitude spectrum consistent with zero, as expected (since
the fitting function is based on the same probability distributions 
as the fast Monte Carlo generator).  
In contrast, the experiment in which the direct and 
cascade decays are generated separately 
shows a small amplitude bias at low and very large \dms. 
Since the bias is small, especially in the region where the limit is set, 
and would cause the limit and sensitivity to be slightly underestimated, 
no attempt is made to correct for this effect;  
instead the deviations of the amplitude from zero observed are 
treated as a systematic uncertainty. 

\end{itemize}

\begin{table}
\tabcaption{Measurement of the \Bs\ oscillation amplitude, ${\cal A}$, for 
various oscillation frequencies together with the statistical uncertainty,
\sigAstat, and the total systematic uncertainty, \sigAsyst; 
a breakdown of  \sigAsyst\
in several categories of systematic effects is also given. }
\labt{syst}
\begin{center}
\begin{tabular}{|l|c|c|c|c|}
\hline
   \dms\               & 0~\invps\    & 5~\invps\  & 10~\invps\   & 15~\invps\
\\
\hline
 ${\cal A}$            &   $-0.030$   & $-0.065$   & 0.303        &  2.291 \\
\rule{0pt}{15 pt}
 \sigAstat             & $\pm 0.099 $ & $\pm0.267$ & $\pm0.590$  & $\pm1.271 $
 \\
\rule{0pt}{15 pt}
 \sigAsyst                      & $^{+0.340}_{-0.340}$ &
  $^{+0.223}_{-0.235}$ & $^{+0.232}_{-0.324}$   & $^{+0.801}_{-0.582}$  \\
\hline
Systematic contributions:     & \ &  \ &  \ &     \\
\rule{0pt}{15 pt}
 --  $R_{\rm b}$, $R_{\rm c}$                 & $^{+0.001}_{-0.001}$  & $^{+0.002}_{-0.001}$ & $^{+0.001}_{-0.002}$ & $^{+0.001}_{-0.005}$  \\
\rule{0pt}{15 pt}
{-- $\fs = \BR{\anti{b}{}{}}{\Bs}$}  &  { $^{+0.046}_{-0.035}$} & 
 { $^{+0.146}_{-0.112}$ } & { $^{+0.133}_{-0.109}$ } & { $^{+0.217}_{-0.173}$ } \\
\rule{0pt}{15 pt}                                                                  
 -- $\fbb = \BR{\particle{b}{}{}}{\mbox{\particle{b}{}{}-baryon}}$ & $^{+0.008}_{-0.010}$  & $^{+0.026}_{-0.018}$ & $^{+0.028}_{-0.023}$ & $^{+0.007}_{-0.002}$  \\
\rule{0pt}{15 pt}
 -- charm fraction             & $^{+0.012}_{-0.012}$  & $^{+0.019}_{-0.016}$ & $^{+0.021}_{-0.018}$ & $^{+0.051}_{-0.043}$  \\
\rule{0pt}{15 pt}
 -- \uds\ fraction             & $^{+0.008}_{-0.008}$ &  $^{+0.023}_{-0.026}$ & $^{+0.032}_{-0.038}$ & $^{+0.078}_{-0.091}$  \\
\rule{0pt}{15 pt}
 -- $ \particle{b}{}{} \rightarrow \ell, 
      \particle{b}{}{} \rightarrow \particle{c}{}{} \rightarrow \ell,
      \particle{b}{}{} \rightarrow \anti{c}{}{} \rightarrow \ell, 
      \particle{c}{}{} \rightarrow \ell$  & $^{+0.065}_{-0.013}$  & $^{+0.000}_{-0.055}$ & $^{+0.000}_{-0.121}$  & $^{+0.464}_{-0.000}$  \\
\rule{0pt}{15 pt}
 -- purities (MC stat.)         & $^{+0.047}_{-0.041}$ &  $^{+0.078}_{-0.070}$ & $^{+0.076}_{-0.075}$ & $^{+0.104}_{-0.108}$  \\
\rule{0pt}{15 pt}
 --  \Bs\ purity classes        & $^{+0.017}_{-0.009}$ &  $^{+0.000}_{-0.007}$ & $^{+0.010}_{-0.018}$ & $^{+0.140}_{-0.187}$  \\
\rule{0pt}{15 pt}
 --  \dmd\                        & $^{+0.037}_{-0.037}$  & $^{+0.002}_{-0.002}$ &  $^{+0.001}_{-0.001}$ & $^{+0.000}_{-0.003}$  \\
\rule{0pt}{15 pt}
 -- \particle{b}{}{}-hadron lifetimes  & $^{+0.033}_{-0.000}$  & $^{+0.000}_{-0.046}$ & $^{+0.027}_{-0.037}$ & $^{+0.282}_{-0.000}$  \\
\rule{0pt}{15 pt}
 -- decay length resolution   & $^{+0.000}_{-0.000}$  & $^{+0.025}_{-0.025}$ &
$^{+0.054}_{-0.057}$ & $^{+0.050}_{-0.021}$  \\
\rule{0pt}{15 pt}
 -- boost term resolution   & $^{+0.010}_{-0.010}$  & $^{+0.030}_{-0.033}$ &
$^{+0.048}_{-0.059}$ & $^{+0.205}_{-0.191}$  \\
\rule{0pt}{15 pt}
 -- \particle{b}{}{}-fragmentation     & $^{+0.023}_{-0.000}$ &  $^{+0.012}_{-0.070}$ & $^{+0.067}_{-0.085}$ & $^{+0.509}_{-0.403}$  \\
\rule{0pt}{15 pt}
{-- \particle{b}{}{}-flavour tagging}  & { $^{+0.317}_{-0.332}$} & 
{  $^{+0.138}_{-0.132}$ }& { $^{+0.132}_{-0.207}$ }&
 { $^{+0.233}_{-0.219}$ } \\
\rule{0pt}{15 pt}
 -- $\DGams/\Gams$ & $^{+0.000}_{-0.002}$ & 
 $^{+0.012}_{-0.000}$ &  $^{+0.011}_{-0.000}$  & $^{+0.018}_{-0.000}$  \\
\rule{0pt}{15 pt}
 -- cascade bias                & $^{+0.060}_{-0.000}$ &  $^{+0.000}_{-0.087}$ &  $^{+0.000}_{-0.085}$  & $^{+0.000}_{-0.069}$  \\
\hline
\end{tabular}
\end{center}
\end{table}

The relative importance of the various systematic uncertainties, as a function of  
\dms, is shown in \Table{syst}. 
Except at low \dms\ the systematic uncertainties are generally small compared to 
the statistical uncertainty. 
At $\Dms=10~\invps$, the most important contributions are from \fs\ and 
the \particle{b}{}{}-flavour tagging.

\section{Checks}
\labs{sec_checks}

Using a fast Monte Carlo generator which takes into account all 
details of the sample composition, the resolution functions, 
the mistag rates and the distributions of $\xeff$, 
the average amplitude over many fast Monte Carlo experiments is
found to be consistent with unity for $\dms= \dmstrue$ and with
zero for any value of \dms\ if $\dmstrue=\infty$.
The estimate, $\sigAstat$, of the statistical uncertainty
on the amplitude has also been verified by studying the distribution of 
${\cal A}/\sigAstat$ for cases where ${\cal A}=0$ is expected.
The mean value and RMS of such a distribution obtained with fast Monte Carlo
experiments generated with $\dmstrue=\infty$ are found to be 
consistent with 0 and 1.

A likelihood fit for \dms\ performed on a \Z{q} Monte Carlo sample having the 
same statistics as the data and generated with a true value of \dms\ of 
3.33~\invps\ yields $\dms = 3.31 \pm 0.12$(stat.)~\invps, in 
agreement with the input value.  
Performing an amplitude fit on the same Monte Carlo events yields the 
results shown in \Fig{dms_mc}. As expected, the amplitude is consistent 
with 1 at the true value of \dms. 
The sensitivity estimated from this Monte Carlo sample
(ignoring systematic uncertainties) is 10.6~\invps, a little higher 
than that obtained from the data, 10.2~\invps,
due to the slightly better decay length 
resolution in Monte Carlo. 

\begin{figure}
\vspace{-2cm}
\begin{center}
\makebox[0cm]{\psfig{file=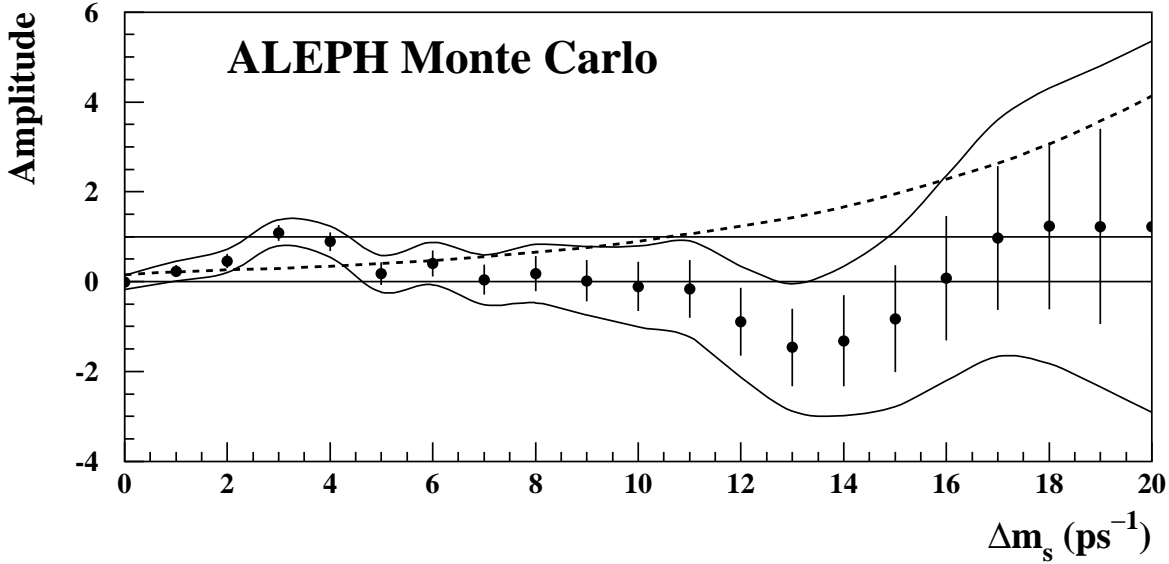,width=1.05\textwidth,
bbllx=0pt,bblly=282pt,bburx=560pt,bbury=560pt}}
\end{center}
\figcaption{Measured \Bs\ oscillation amplitude as a function of \dms\ in the \Z{q} Monte Carlo.
The error bars represent the 1$\sigma$ statistical uncertainties, 
the solid curve the one-sided \CL{95} contour (systematic effects
not included). The dotted line is $1.645\,\sigma$. The generated value of \dms\ 
is 3.33~\invps. }
\labf{dms_mc}
\end{figure}

As a further check of the assumed mistags and sample composition, 
the analysis is used to measure \dmd\ in the data. 
Fixing \dms\ to 50~\invps\ and minimizing the 
negative log-likelihood with respect to 
\dmd\ gives $\dmd = 0.451 \pm 0.024$(stat.)~\invps,
consistent with the latest world average of $0.463 \pm
0.018~\invps$~\cite{Schneider}. 
\Figure{dmd_amp} shows that the 
fitted \Bd\ oscillation amplitude is consistent with that observed in 
the \Z{q} Monte Carlo and has the expected value of 1 at the 
minimum of the negative log-likelihood. 
To check that the sample composition and mistags assumed for each 
\Bs\ purity class and tagging class are reasonable, a
fit for the \Bd\ oscillation amplitude is performed separately 
in each class. 
At $\dmd = 0.451~\invps$ a value of ${\cal A}$ consistent with 1 is found in 
all classes; the largest deviation being 
$1.5\,\sigma_{\mbox{\scriptsize stat}}$ in the last \Bs\ purity  
class (``remainder'').  

\begin{figure}
\vspace{-2cm}
\begin{center}
\makebox[0cm]{\psfig{file=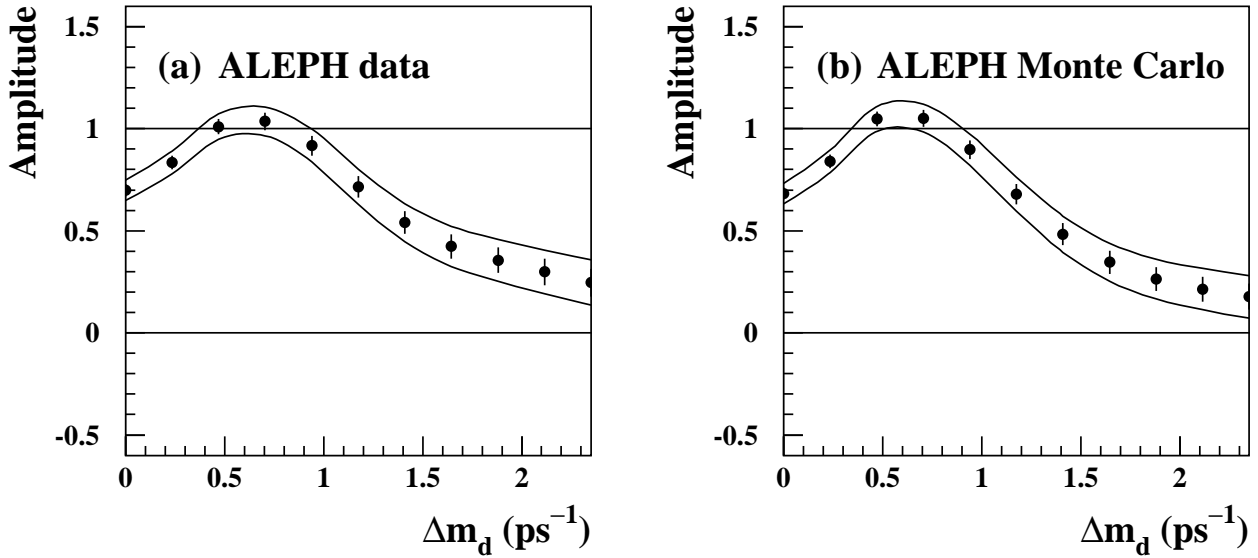,width=1.05\textwidth,
bbllx=0pt,bblly=285pt,bburx=560pt,bbury=560pt}}
\end{center}
\figcaption{Measured \Bd\ oscillation amplitude as a function of \dmd\ in (a) the data and (b)
the \Z{q} Monte Carlo.
The error bars represent the 1$\sigma$ total uncertainties and
the curves the one-sided \CL{95} contour (systematic effects not included).}
\labf{dmd_amp}
\end{figure}

\section{\boldmath Combination with \Ds\ analyses} \labs{combination}
\begin{figure}
\vspace{-1.7cm}
\begin{center}
\makebox[0cm]{\psfig{file=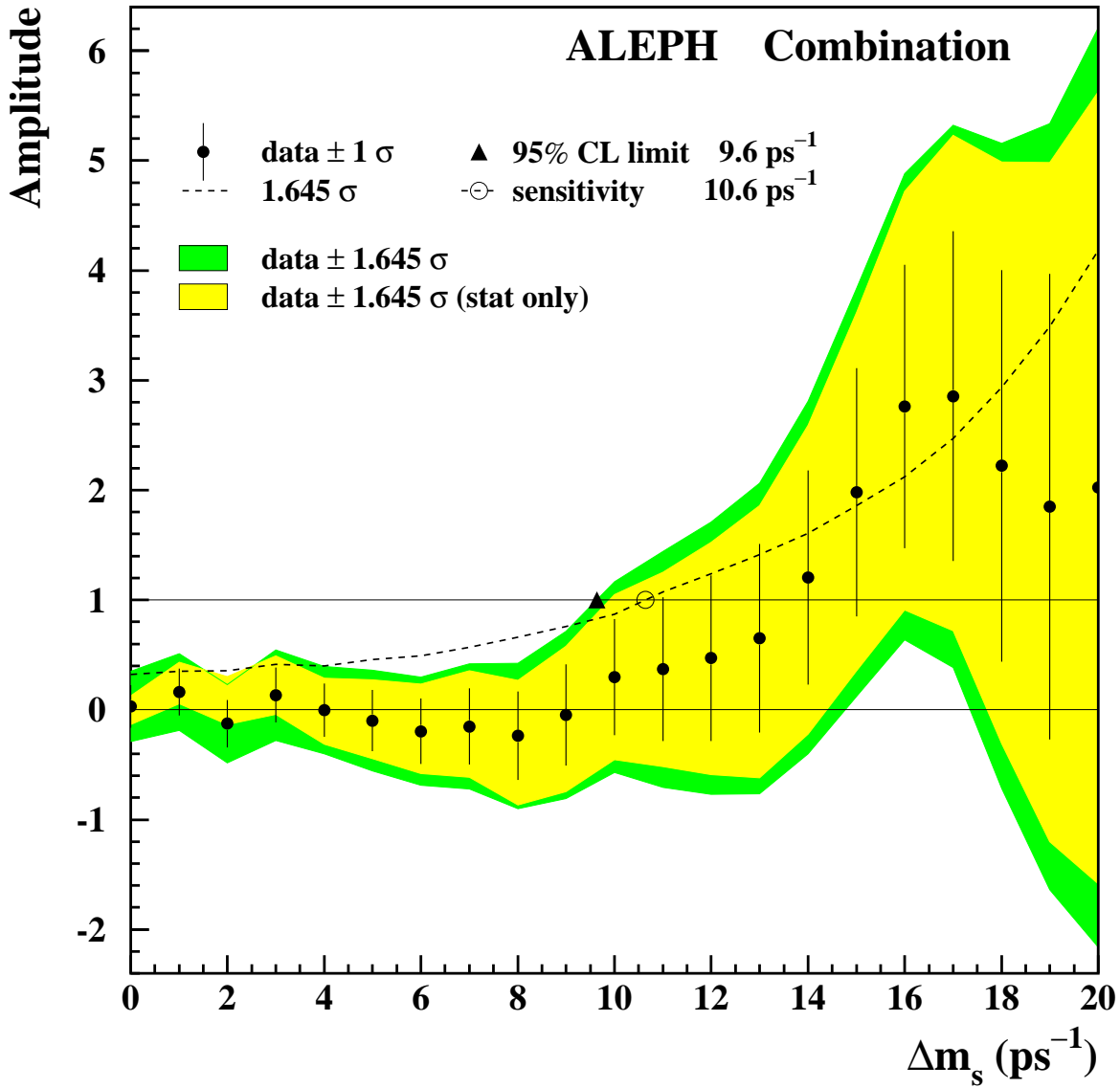,width=1.05\textwidth}}
\end{center}
\vspace{-0.3cm}
\figcaption{Measured \Bs\ oscillation amplitude as a function of \dms\ for the 
combination of this analysis with the ALEPH \Ds\ based analyses.}
\labf{amp_comb}
\end{figure}

The amplitudes measured in this 
analysis and in the two ALEPH \Ds\ analyses~\cite{ALEPH-DS-LEPTON,ALEPH-DSHAD}
are combined.
The small number of events common to both this analysis and the \Ds--lepton 
analysis are removed from the inclusive lepton sample before combining the results.
The following sources of systematic uncertainty are treated as fully 
correlated:  
the values assumed for \fs, \fbb, \dmd\ and the various \particle{b}{}{}-hadron lifetimes,
the \particle{b}{}{} fragmentation,
the decay length resolution bias in the Monte Carlo simulation $\Sldat$ and $\fldat$,
the mistag probabilities, and the use of the effective discriminating variable.
Since the physics parameters assumed in the three analyses are slightly 
different, the \Ds\ results are adjusted to the more recent 
set of physics parameters listed in \Table{phyparams} before averaging.
The combined amplitude plot is displayed in \Fig{amp_comb} and the corresponding 
numerical values are listed in \Table{amplitude_combined}.
All values of \dms\ below 9.6~\invps\ are excluded at \CL{95}. 
The combined sensitivity is 10.6~\invps.

As the statistical correlation between this analysis and the previous
ALEPH dilepton and lepton-kaon analyses \cite{ALEPH-DILEPTON, ALEPH-WARSAW-COMBINATION}
is very large, no significant improvement in sensitivity is expected
if these latter analyses were included in the combination.

\begin{table}
\tabcaption{Combined measurements of the \Bs\ oscillation amplitude ${\cal A}$ 
 as a function of \dms\ (in \invps), 
together with the statistical uncertainty $\sigAstat$ and
the total systematic uncertainty $\sigAsyst$.}
\labt{amplitude_combined}
\begin{center}
\begin{tabular}{|c|r@{$\,\pm$}r@{$\,\pm$}r|c|c|r@{$\,\pm$}r@{$\,\pm$}r|c|c|r@{$\,\pm$}r@{$\,\pm$}r|}
\cline{1-4} \cline{6-9} \cline{11-14} 
\dms\ & $\rule[-5pt]{0pt}{18pt} {\cal A}$ & $\sigAstat$ & $\sigAsyst$ & &
\dms\ & $\rule[-5pt]{0pt}{18pt} {\cal A}$ & $\sigAstat$ & $\sigAsyst$ & &
\dms\ & $\rule[-5pt]{0pt}{18pt} {\cal A}$ & $\sigAstat$ & $\sigAsyst$ \\
\cline{1-4} \cline{6-9} \cline{11-14} \\[-12pt]
\cline{1-4} \cline{6-9} \cline{11-14} 
$ 0.00$ & $+0.03$ & $ 0.08$ & $ 0.18$ & & $ 7.00$ & $-0.15$ & $ 0.30$ & $ 0.18$ & & $14.00$ & $+1.21$ & $ 0.86$ & $ 0.47$ \\ 
$ 1.00$ & $+0.16$ & $ 0.11$ & $ 0.18$ & & $ 8.00$ & $-0.24$ & $ 0.35$ & $ 0.21$ & & $15.00$ & $+1.98$ & $ 0.99$ & $ 0.54$ \\ 
$ 2.00$ & $-0.13$ & $ 0.13$ & $ 0.17$ & & $ 9.00$ & $-0.05$ & $ 0.40$ & $ 0.23$ & & $16.00$ & $+2.76$ & $ 1.16$ & $ 0.57$ \\ 
$ 3.00$ & $+0.13$ & $ 0.16$ & $ 0.19$ & & $10.00$ & $+0.30$ & $ 0.46$ & $ 0.27$ & & $17.00$ & $+2.86$ & $ 1.37$ & $ 0.61$ \\ 
$ 4.00$ & $+0.00$ & $ 0.18$ & $ 0.16$ & & $11.00$ & $+0.37$ & $ 0.54$ & $ 0.37$ & & $18.00$ & $+2.22$ & $ 1.61$ & $ 0.77$ \\ 
$ 5.00$ & $-0.10$ & $ 0.22$ & $ 0.18$ & & $12.00$ & $+0.47$ & $ 0.64$ & $ 0.39$ & & $19.00$ & $+1.85$ & $ 1.88$ & $ 0.98$ \\ 
$ 6.00$ & $-0.20$ & $ 0.25$ & $ 0.17$ & & $13.00$ & $+0.65$ & $ 0.75$ & $ 0.42$ & & $20.00$ & $+2.02$ & $ 2.19$ & $ 1.29$ \\ 
\cline{1-4} \cline{6-9} \cline{11-14} 
\end{tabular}
\end{center}
\end{table}

\section{Conclusion}

From a sample of 33023 inclusive lepton events, 
all values of \dms\ below 9.5~\invps\ are excluded 
at \CL{95} using the amplitude method. This analysis supersedes the previous
ALEPH inclusive lepton analysis \cite{ALEPH-LEPTON-JET-WISCONSIN} and provides the 
highest sensitivity and highest \CL{95} lower limit on \dms\ of any \Bs\ mixing 
analysis published to date~\cite{ALEPH-DS-LEPTON,ALEPH-DSHAD,ALEPH-DILEPTON,
ALEPH-LEPTON-JET-WISCONSIN,DELPHI-DMS-COMBINATION,OPALDMS}. 

Taking into account correlated systematic uncertainties the combination 
with the ALEPH \Ds\ based analyses yields $\dms>9.6~\invps$ at \CL{95}.

\section*{Acknowledgements}
It is a pleasure to thank our colleagues in the accelerator divisions of CERN
for the excellent performance of LEP.
Thanks are also due to the technical
personnel of the collaborating institutions for their support in constructing
and maintaining the ALEPH experiment. Those of us not from member
states wish to thank CERN for its hospitality.

\end{document}